\RequirePackage{ifpdf}
\documentclass{PoS}

\newcommand{\skipthis}[1]{}

\newcommand\noi{\noindent}

\newcommand{\EoR}{{\sl Epoch of Reionization}}
\newcommand{\CD}{{\sl Cosmic Dawn}}
\newcommand{\DA}{{\sl Dark Ages}}
\newcommand{\SKA}{{\sl Square Kilometre Array}}
\newcommand{\CMB}{{\sl Cosmic Microwave Background}}
\newcommand{\WF}{{\sl Wouthuysen-Field}}
\newcommand{\LyA}{Lyman-$\alpha$}
\newcommand{\JWST}{{\sl James Webb Space Telescope}}

\usepackage{graphicx}

\usepackage{natbib}

\title{The Cosmic Dawn and Epoch of Reionization with the Square Kilometre
Array}

\ShortTitle{CD/EoR with SKA}

\author{
\speaker{L.V.E. Koopmans}\thanks{Chair CD/EoR SWG},$^1$
J.~Pritchard\thanks{Vice-Chair CD/EoR SWG}, 
G.~Mellema\thanks{Vice-Chair CD/EoR SWG}, 
F.~Abdalla, 
J.~Aguirre, 
K.~Ahn, 
R.~Barkana, 
I.~van Bemmel,
G.~Bernardi, 
A.~Bonaldi, 
F.~Briggs,
A.G.~de Bruyn, 
T.C.~Chang, 
E.~Chapman,
X.~Chen, 
B.~Ciardi, 
K.K.~Datta, 
P.~Dayal, 
A.~Ferrara, 
A.~Fialkov,
F.~Fiore, 
K.~Ichiki, 
I.~T.~Illiev, 
S.~Inoue, 
V.~Jeli\'{c}, 
M.~Jones,
J.~Lazio, 
U.~Maio, 
S.~Majumdar, 
K.~J.~Mack, 
A.~Mesinger, 
M.F.~Morales,
A.~Parsons, 
U.-L.~Pen, 
M.~Santos, 
R.~Schneider, 
B.~Semelin, 
R.S.~de Souza,
R.~Subrahmanyan, 
T.~Takeuchi, 
C.~Trott, 
H.~Vedantham, 
J.~Wagg,
R.~Webster, 
S.~Wyithe
\\
$^1$Kapteyn Astronomical Institute, University of Groningen, The Netherlands\\
Other author institutions are listed in the appendix\\
E-mail: \email{koopmans@astro.rug.nl}}

\abstract{ Concerted effort is currently ongoing to open up the \EoR\
($z\sim$15-6) for studies with IR and radio telescopes.  Whereas IR
detections have been made of sources (\LyA\ emitters, quasars and
drop-outs) in this redshift regime in relatively small fields of view, no
direct detection of neutral hydrogen, via the redshifted 21-cm line, has
yet been established. Such a direct detection is expected in the coming
years, with ongoing surveys, and could open up the entire universe from
$z\sim$6-200 for astrophysical and cosmological studies, opening not only
the \EoR, but also its preceding \CD\ ($z\sim$30-15) and possibly even the
later phases of the \DA\ ($z\sim$200-30). All currently ongoing experiments
attempt statistical detections of the 21-cm signal during the \EoR, with
limited signal-to-noise. Direct imaging, except maybe on the largest
(degree) scales at lower redshifts, as well as higher redshifts will remain
out of reach. The \SKA\ (SKA) will revolutionize the field, allowing direct
imaging of neutral hydrogen from scales of arc-minutes to degrees over most
of the redshift range $z\sim$6-28 with SKA1-LOW, and possibly even higher
redshifts with the SKA2-LOW. In this SKA will be unique, and in parallel
provide enormous potential of {\it synergy} with other upcoming facilities
(e.g. JWST). In this chapter we summarize the physics of 21-cm emission,
the different phases the universe is thought to go through, and the
observables that the SKA can probe, referring where needed to detailed
chapters in this volume. This is done within the framework of the current
SKA1 baseline design and a nominal CD/EoR straw-man survey, consisting of a
shallow, medium-deep and deep survey, the latter probing down to
$\sim$1\,mK brightness temperature on arc-minute scales at the end of
reionization. Possible minor modifications to the design of SKA1 and the
upgrade to SKA2 are discussed, in addition to science that could be done
already during roll-out when SKA1 still has limited capabilities and/or
core collecting area.  }

\FullConference{
Advancing Astrophysics with the Square Kilometre Array\\
June 8-13, 2014\\
Giardini Naxos, Italy}

\begin{document}

\section{Context and Layout}


\noi In this review chapter the impact of the \SKA\ (hereafter SKA1 and
SKA2 for the first and second phases of construction) in the field of
high-redshift 21-cm observations of neutral hydrogen during the EoR and CD
is outlined, building partly on the white paper of
\citet[][]{2013ExA....36..235M} and referring to more than a dozen related
science chapters in this volume that discuss particular aspects  in much
greater detail. In Section~\ref{sect:overview} an overview of processes
occurring during the \DA, \CD\ and \EoR\ (hereafter DA/CD/EoR) eras are
given. In Section~\ref{sect:physics} the main physical processes that take
place during these eras are discussed, with reference to the relevant
accompanying chapters in this volume. In Section~\ref{sect:observables}
observables of the redshifted 21-cm line are discussed, whereas in
Section~\ref{sect:surveydesign} we discuss the relevant parameters for any
21-cm survey design. In Section\,\ref{sect:survey}, we discuss a
three-tiered survey with the SKA of the redshifted 21-cm emission from the
CD/EoR.  In Section\,\ref{sect:rollout} we discuss what can be done with a
half (50\% collecting area) and a full SKA1 and with SKA2 (nominally 4
times SKA1), during the rollout and build-out of SKA1 to SKA2.  In
Section\,\ref{sect:end} we draw some general conclusions and provide
suggestions. Throughout this chapter when referring to SKA1/2, we mean
SKA1/2-LOW.

\section{A Short History of Neutral Hydrogen from Recombination to Reionization}\label{sect:overview}

\begin{figure}
\includegraphics[width=0.99\textwidth]{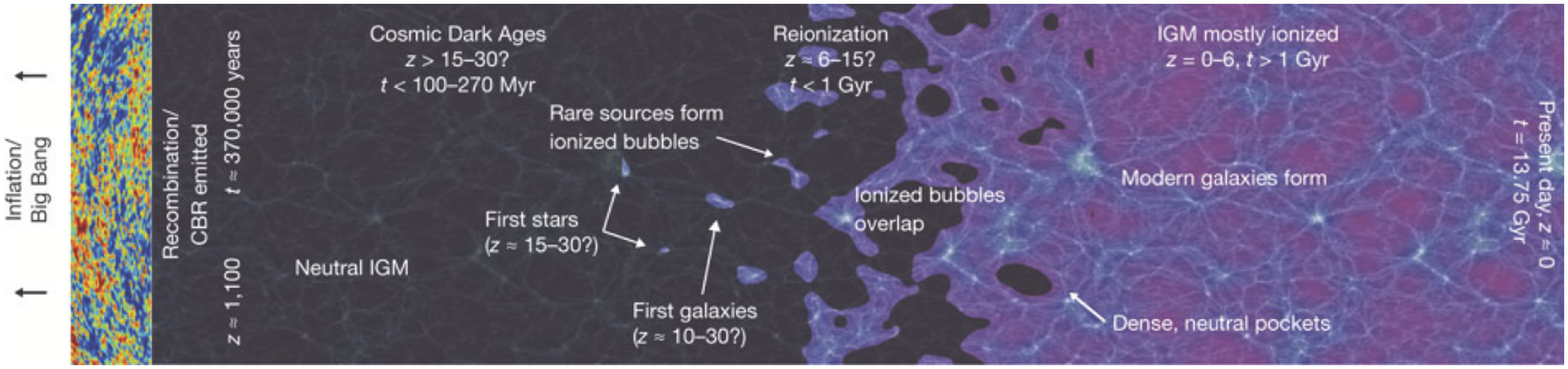}
\caption{A graphical overview of the main phases of neutral and ionized hydrogen in the Universe. From \citet[][]{2010Natur.468...49R}.}
\label{fig1}\label{fig:EoR}
\end{figure}

\noi This section shortly summarizes the main events that occurred in the Universe between recombination and reionization (see Fig.\ref{fig:EoR}),
focussing on those aspects that are relevant to interferometric measurements of the 21-cm hyperfine transition line of 
neutral hydrogen with the SKA. We refer to \citet[][]{2001PhR...349..125B, 2004NewAR..48.1039F, 2007PhRvD..76h3005L, 2008PhRvD..78j3511P, 2010ARA&A..48..127M, 2013fgu..book.....L, 2014MNRAS.437L..36F} for many more details.

\subsection{Emission and absorption at 21-cm from z$\sim$1100 to z$\sim$6}

\noi Just after recombination at $z\sim 1100$, the spin temperature of
neutral hydrogen coupled very effectively to the neutral gas temperature,
which itself coupled to the \CMB\ (CMB hereafter) photons via Compton
heating because of a small trace density of electrons, making it impossible
to observe 21-cm radiation either in emission or absorption.  This period
can truly be regarded as an "Age of Ignorance" in cosmology with no known
directly observable tracer of either dark matter or baryons. Around $z\sim 200$,
the neutral gas temperature  decoupled from the CMB and adiabatically
cooled in an expanding Universe.  During this period (lasting to $z\sim
30$), called the \DA,  the 21-cm line can be seen in absorption and the
physics of this period is relatively well-understood.  At the end of the
\DA, the spin-temperature coupling to the CMB started to dominate over its
coupling to the cold gas, increasing the spin-temperature such that its
differential brightness temperature approached zero once more. Around the
same time, at $z\sim 30$, haloes of $\sim 10^6$ solar masses started to
form in sufficiently large numbers.  Gas could cool sufficiently in to
their potential wells to form the first (Pop-III) stars. These stars both
radiated and heated their surrounding gas. Via the \WF\ (W-F) effect, the
spin-temperature was able to couple to the cold gas temperature, creating a
second period (the \CD) where the 21-cm line was seen in absorption. While
star-formation continued, it is thought that the first X-ray emitting
sources appeared, heating the neutral gas and thereby (again via the \WF\
effect) raising the overall spin-temperature to a level at or above the CMB
temperature. Consequently, the 21-cm line became visible in emission around
$z\sim 15$. While radiating and heating, these first sources (possibly
including mini-quasars) also ionized the gas around them and a period of
{\sl Reionization} started that is thought to have lasted until $z \sim
5-6$. During the latter phase most of the neutral hydrogen became ionized
and is thought to still reside in the IGM today. During the same period the
gas was metal enriched and seed black-holes  grew to super-massive black
holes already seen at redshifts as high as $z\sim 7$. As discussed in the
accompanying chapters in this volume, observing the \DA, \CD\ and \EoR\ is
crucial for our understanding of the Universe which we see from the present
day up to high redshifts and provide an enormous potential for synergy with
other IR to sub-mm facilities (e.g.\  Euclid, ELT/TMT/GMT, JWST, ALMA,
etc). As such, their study features high on nearly all future science and
instrumental roadmaps, including that of the SKA, and is one of the main
science drivers for a range of SKA precursors and pathfinders (e.g.\ LOFAR,
MWA, PAPER and GMRT, and the planned
HERA\footnote{http://reionization.org/}). 

\subsection{Current constraints on the Cosmic Dawn and Epoch of Reionization}\label{sect:currentobs}

\noi Even though the history of neutral hydrogen is crucial to our understanding of the high-redshift Universe, very little is known about
it. There is only a handful of observations that currently constrains physical models of the evolution of neutral hydrogen: 
\begin{itemize} 

\item the Gunn-Peterson trough at/near the end of reionization from which
(a limit on) the HI optical depth can be inferred
\citep[e.g.][]{2001AJ....122.2850B, 2002AJ....123.1247F,
2006AJ....132..117F}, 

\item  the discovery of Gamma-Ray Bursts (GRB) during and after
reionization \citep[e.g.][]{2007ApJ...669....1R, 2014ApJ...785..150S} from
which the high-mass star-formation rate (SFR) can be inferred,

\item the metal abundances and old stellar populations
\citep[e.g.][]{2012Natur.492...79S} possibly in Damped Lyman Absorbers
(DLAs),

\item the evolution of the IGM temperature after reionization from which a
nominal reionization redshift can be inferred, if the IGM cooled
adiabatically and did not re-heat \citep[e.g.][]{2002MNRAS.332..367T,
2010MNRAS.406..612B}, 

\item the polarization of the CMB due to ionized gas (basically
electrons) integrated along the line-of-sight to the present day from which
a median reionization redshift and duration can be inferred \citep[see
e.g.][]{2015ApJ...802L..19R}, 

\item the integrated IR and X-ray backgrounds setting limits on the emission from the first (Population III \& II) stars, X-ray sources and quasars \citep[e.g.][]{2014Sci...346..732Z, 2013MNRAS.433.2047F, 2014ApJ...785...38H}

\item the observations of the first galaxies through drop-out techniques, currently out to $z\sim 12$ already \citep[e.g.][]{2013ApJ...763L...7E}, or \LyA\ emission \citep[e.g.][]{2014MNRAS.440.2375M}, providing limits on ionizing (stellar) sources, and the HI content in the IGM from \LyA\ emitters. 

\end{itemize}

\noi Although theoretical models are highly degenerate 
especially at increasingly larger redshifts, it does appear from these observations that reionization is roughly half-way at $z\sim 10$,
followed by a rapid increase in the SFR-density below this redshift \citep[e.g.][]{2011MNRAS.414.1455L, 2012ApJ...752L...5B, 2013ApJ...763L...7E}. Around $z\sim 5-7$ potentially 10\% of 
hydrogen could still be in neutral patches \citep[e.g.][]{2010MNRAS.407.1328M}, although most of it was (re)ionized by then, making most of the the universe 
transparent to {\sl uv}-radiation
all the way to the present day. Beyond $z \sim 10$ even less is known and most of what we think was happening is
actually based on theory (e.g.\ Pop-III stars) and simulations that are constrained by physical models extrapolated from lower redshifts 
\citep[e.g.\ X-ray heating via XRBs; ][]{2014Natur.506..197F}. It it therefore crucial that new observational avenues are explored to open up these early phases in the Universe for observational studies. 
The redshifted 21-cm emission of 
neutral hydrogen provides just such a possibility. Neutral hydrogen is all-pervasive until its (re)ionization completes and it complements (i.e.\ anti-correlates)
with sources of ionization, with the CMB and with the IR/X-ray background \citep[although see][]{2015PhRvL.114j1303F}. Moreover due to the line-nature of the 21-cm
hyperfine transition, redshift (and velocity) information can be gained directly from 21-cm emission by fine-tuning observations to different radio 
frequencies [i.e. to $1420/(1+z)$\,MHz]. Hence tomography of (redshifted) 21-cm emission can be done by covering a wide range of frequencies. 

\noi In the case of
the SKA, frequencies of 50\,MHz (during phase 1) and higher allow HI mapping at redshifts below $z\sim27$ ($\sim\,120$ Myr after the Big Bang) all the
way to the end of reionization $z\sim5-6$ ($\sim 1$ Gyr after the Big Bang). 
Especially the highest redshifts are extremely hard to reach with other instruments, even with the \JWST\ (JWST) in the next decade.

\subsection{Current 21-cm detection experiments}

\noi There are two general approaches to observe HI emission via the 21-cm line: (i) its spatially-averaged global signal and (ii) its spatial fluctuations, both as function of frequency (i.e.\ redshift, time, distance). In the former case
the average redshifted 21-cm emission\footnote{More precisely its brightness temperature.} at a given redshift is measured over a very large (possibly global) area of the sky such that spatial fluctuations are averaged away, in general
via single dipoles or fully-filled antennae (e.g.\ dish, aperture arrays with near unity filling factor). The global signal can vary between a few mK brightness temperature near the end of reionization and around those redshift where the HI spin temperature and the CMB were
almost same, to as much as +30\,mK when the spin temperature was well above the CMB temperature, and to as low as $-$100\,mK or even 
less during the \CD\ when HI was seen in absorption against the CMB and the spin-temperature coupled to the cold neutral gas. 
Whereas in principle these signals could be detected in a few hours to a few days based purely on thermal noise limits, such
a detection turns out to be extremely difficult to the high dynamic range that is needed in band-pass calibration (i.e.\ $>10^6$), in the presence of RFI
and the complex spatially-temporal-frequency varying sky that couples to a generally polarized and frequency-dependent receiver beam.
Whereas many global experiments are ongoing (e.g. EDGES, CORES, SARAS,
LEDA, etc; see Subrahmanyan et al. 2015) none have reached the few to tens of mK level yet required for a solid detection although bandpass stability of $\sim 1$ K has now been reached \citep[][]{2013ExA....36..319P, 2015ApJ...801..138P}.  

An alternative/complementary approach to detect the redshifted 21-cm emission is via an interferometer, which is insensitive to the global signal (which is detectable on the (near)zero spacings only), but can only detect fluctuations in the 21-cm brightness temperature. Although these fluctuations can be considerable, they are 
far smaller in amplitude than the global signal hence requiring long ($\sim$1000\,hrs) of integration time to detect. At lower redshifts during the \EoR, however,
ionized bubbles could be $-30$\,mK in depth and detectable on somewhat longer baselines with arcmin resolution. Apart from the
ionized bubbles, brightness temperature fluctuations are sourced
by density fluctuations in the time-evolving dark-matter distribution and via the spin-temperature coupling to the neutral gas and/or the 
CMB, though the \WF\ effect and through heating. In addition, fluctuations are sourced by peculiar and bulk-flow velocities
that Doppler-shift the 21-cm line, affecting its brightness temperature. Whereas this complex astrophysics make interpretation of the 
21-cm signal harder, it also provides a wealth of information on sources and physical process occurring during that period. In combination with the global signal and
other observables (e.g.\ galaxies, CMB polarization, etc)  these brightness-temperature fluctuations provide a treasure trove of information. The second advantage of using an interferometer
is to reduce the levels of foreground emission from the Milky Way (several Kelvin versus several hundred Kelvin) that could potentially contaminate
the feeble 21-cm signal, as well as the fact that interferometers are often easier to calibrate than total-power experiments. At the moment each of
these approaches are pursued by multiple teams and valuable in their own right and targeting different parameter spaces, thereby complementing each other. Current HI detection experiments are ongoing with the GMRT \citep[e.g.][]{2011MNRAS.413.1174P},
LOFAR \citep[e.g.][]{2013A&A...550A.136Y, 2013A&A...556A...2V}, PAPER \citep[e.g.][]{2015arXiv150206016A, 2015arXiv150300045P}, 
MWA \citep[e.g.][]{2014PhRvD..89b3002D} and observations with the LWA (i.e.\ LEDA) \citep[e.g.][]{2014era..conf10301G},
as well as with NenuFAR\footnote{http://nenufar.obs-nancay.fr/Argumentaire-scientifique.html}, under construction, are planned.  

\subsection{Planned high-redshift 21-cm arrays} 

\noi Despite great strides forward over the last decade, no high-redshift 21-cm detection has yet been successfully claimed, neither of the global HI signal nor of its fluctuations. 
Even if successful, all current observational programs (except for those targeting the global signal) aim for a {\sl statistical} detection via power-spectra (and high-order statistics)
and only on the very largest angular scales (degrees) could one potentially reach a S/N$\sim$few in a few MHz frequency bin, comparable to 
the first CMB maps with COBE \citep[][]{1992ApJ...396L...1S}. To overcome the current S/N$\sim$1 barrier substantially more collecting area is required, especially on short baselines corresponding
to a few to tens of arc minute scales, by an order of magnitude over the current LOFAR-core collecting area, the latter being the largest low frequency 
array currently aiming to detect high-$z$ 21-cm emission. Such a collecting area is 
foreseen in the current baseline design of SKA1 and might evolve in to SKA2 with four times the SKA1 collecting area (see this volume), allowing one to move from upcoming statistical detections to tomographic (i.e.\ direct imaging) measurements of HI in the next decade. A similar US-lead effort called HERA\footnote{http://reionization.org/papers/} is also planned extending the current PAPER project in South-Africa, and NenuFAR in France that aims for a nearly fully filled low-frequency (10-80MHz) array of 400-m diameter.

\section{Some relevant physics during the CD/EoR eras}\label{sect:physics}

\noi Radio-telescopes measure the sky's intensity distribution (more
precisely its Fourier transform), which traditionally is expressed in terms
of a brightness temperature $T_{\rm b}$ in the Rayleigh-Jeans regime. It
can be shown that the brightness temperature of neutral hydrogen as
function of redshift z, seen against the CMB, can be written as
\citep[e.g.][]{1997ApJ...475..429M}:

\begin{eqnarray}
\delta T_{\rm b} = 27 x_{\rm HI} (1 + \delta) \sqrt{\frac{1+z}{10}}\left(\frac{T_{\rm s}-T_{\rm CMB}}{T_{\rm s}} \right) \left(\frac{\Omega_{\rm b}}{0.044}\frac{h}{0.7}\right) \sqrt{\frac{\Omega_m}{0.27}} \nonumber \\
\times \left(\frac{1-Y_{\rm p}}{1-0.248}\right) \left(1 +
\frac{1}{H(z)}\frac{dv_{||}}{dr_{||}}\right)^{-1} {\rm ~~~mK} .
\label{eqn:dt}
\end{eqnarray}

This equation contains several terms set by cosmology (i.e.\ the global baryonic density and spatial density fluctuations, 
$\Omega_{\rm b}$ and $\delta$, respectively; the total mass density and the Hubble constant $\Omega_{\rm m}$ and $H(z)$ or $h$, respectively), by (g)astro-physics (i.e.\ the spin-temperature $T_{\rm s}$ of HI, the CMB temperature $T_{\rm CMB}$) and by Doppler effects ($dv_{||}/dr_{||}$; e.g.\ due to peculiar motions and bulk-flows of the gas). Via the brightness temperature and different measures of it, e.g.\ variance, higher-order statistics, power-spectra, $n$-point correlations, tomographic cubes (i.e. images), HI-absorption spectra, cross-correlation, etc., one can address questions about the physics and sources responsible for processes during the CD/EoR such as heating (changing $T_{\rm gas}$), \LyA\ emission and the Wouthuysen-Field effect (changing $T_{\rm s}$ via $T_{\rm gas}$; \citet[][]{1952AJ.....57R..31W, 1958PIRE...46..240F, 1959ApJ...129..536F}), reionization (changing $x_{\rm HI}$) and the growth of density fluctuations and peculiar velocities (changing $\delta$ and $dv_{||}/dr_{||}$).

\subsection{Phases of 21-cm emission and absorption}

\noi Largely due to the expansion of the universe and within it the gravitational collapse of matter, the universe at some point becomes
non-linear in its (evolving) density structures producing the first stars, stellar remnants, black holes, and galaxies. These objects
in turn will impact the IGM/HI surrounding them, leading to a host of observational effects that can be used to assess their 
properties. Below we shortly summarize the main phases of HI after recombination ($z\sim 1100$) and before complete 
reionization ($z \sim 5-6$). We again refer to \citet[][]{2001PhR...349..125B, 2004NewAR..48.1039F, 2007PhRvD..76h3005L, 2008PhRvD..78j3511P, 2010ARA&A..48..127M, 2013fgu..book.....L, 2014MNRAS.437L..36F} for many more details. 

\begin{itemize}

\item{\bf Age of Ignorance  (z$\sim$1100-200):}  Just after recombination the spin-temperature of neutral hydrogen tightly coupled
to the CMB temperature, indirectly via the gas temperature (via Compton scattering), making the 21-cm line invisible against the 
CMB temperature. At the moment no known observational technique
is able to probe this era, even if it were possible at these high redshifts. Observing these high redshift using the 21-cm line would first
of all have to be carried out in space to avoid the ionospheric plasma frequency, but most likely would be limited by self-absorption in the MW \citep[][]{1978ApJ...221..114N}.
In case of dark-matter annihilation \citep[see e.g.][]{2013MNRAS.429.1705V} or other heating sources (\LyA\ photons from recombination; \citep[see][]{2013JCAP...11..066F}) could one conceivably have the spin-temperature decouple from the CMB
temperature. However, it is likely that this period in the history of the Universe will remain out of reach till far in the future, and remain
a true {\sl Age of Ignorance}.

\item{\bf Dark Ages (z$\sim$200-30): } About five million years after recombination the spin-temperature of hydrogen coupled -- via trace electrons -- to the colder-than-the-CMB gas-temperature. Neutral hydrogen became visible in absorption again the CMB during that era. At the end of the \DA\ the density of trace electrons, however, dropped sufficiently (through expansion of the Universe) to make the coupling of the spin and gas temperature inefficient. The spin temperature started to follow the CMB temperature again and $T_{\rm b}$ approached zero from below. However, this phase just preceded the formation of the first radiating sources and probably only lasted  briefly or might not even have been fully reached (i.e.\ $T_{\rm b}\sim 0$\,mK). {\sl Processes that can be studied through measurements of $T_{\rm b}$ during the \DA\ are the dark-matter power-spectrum evolution and its annihilation physics, baryonic bulk-flows \citep[][]{2010PhRvD..82h3520T} and the physics of gravity and general relativity.} The physics during this era can largely be understood through linear theory \citep[][]{2007PhRvD..76h3005L}, or linear corrections
thereof \citep[e.g.][]{2014PhRvD..89h3506A}
and deviations of observation from CDM predictions will immediately indicate new physics. Studying HI deep in to the \DA\ however will remain out of reach in the near future and requires space-based radio telescopes because 21-cm emission will have been redshifted to near or below the ionospheric plasma frequency.

\item{\bf Cosmic Dawn (z$\sim$30-15):} The \CD\ is typically defined as the
time when the first stars (or other radiating sources) were formed. \LyA\
emission from these sources efficiently coupled the spin temperature to
that of the cold gas (via the Wouthuysen-Field effect), again leading to
neutral hydrogen seen in absorption. At the same time, however, the gas
itself was heated supposedly via X-ray heating or, possibly, via the hot
ISM and \LyA\ photons (Pacucci et al.\ 2014). This gas-heating and the
parallel process of coupling the spin and gas temperature led to a rapid
rise in the brightness temperature of neutral hydrogen until it is finally
seen in emission around $z\sim 15$. One should note that many of these
processes are still ill-understood and all these effects (heating,
coupling, etc) could shift around in redshift substantially, especially
between the \CD\ and the \EoR.  Hence when these processes exactly occurred
is not known and redshifts indicated here are merely indicative numbers
currently expected from nominal models. {\it Processes that can be studied
during the CD are the formation of the first (pop-III) stars, the first
BHs, X-ray heating sources, W-F coupling, bulk-flows, etc.}

\item{\bf Epoch of Reionization (z$\sim$15-6):} While heating together with
the W-F effect change the spin-temperature of the neutral hydrogen, the
same radiation field (i.e. {\sl uv}-radiation) starts to ionize hydrogen
leading to the percolation of bubbles around the first mini-haloes
containing (Pop-III) stars and possible intermediate mass black holes,
i.e.\ mini-quasars. As time progresses and the universe becomes
increasingly more non-linear (on small scales), more stars and quasars are
formed.  Although recombination can have some impact \citep[see
e.g.][]{2014MNRAS.440.1662S}, it can not stop or balance ionization in an
expanding universe and by $z \sim 6$ the entire universe, apart from
pockets of neutral hydrogen (mostly in galaxies), will be ionized once
more. {\it Processes that can be studied during the \EoR\ are the physics
of the ionizing sources, such as pop-III and II stars, mini-quasars,
feedback to the IGM and the transition to the currently visible universe.}

\item{\bf Post-Reionization (z<6):} Whereas most neutral hydrogen will have been ionized, pockets of neutral hydrogen might remain even at lower redshifts and in galaxies \citep[e.g.][]{2010MNRAS.407.1328M}, which could be studied though intensity mapping and via cross-correlations with other e.g. IR surveys (see this volume).

\end{itemize}

\subsection{Sources of heating, radiation, feedback and ionization}\label{sect:sources}

\noi Whereas the above situation  delineates general eras during which certain physical processes (e.g. \LyA\ emission,
heating, ionization, etc) might dominate, the understanding of these processes is far from certain and depends strongly on the
types of sources responsible for {\sl uv}- and X-ray emission. Via their imprint on the brightness temperature of neutral hydrogen one can 
gain insight in to these first radiating sources, which is one of the main reasons to study high-$z$ 21-cm emission.
Below a range of sources is listed that are currently thought to play a role. We refer to the accompanying chapter in this volume for more details.

\begin{itemize}

\item{\bf Population-III and -II Stars} 

The first stars presumably formed just after the \DA\ from relatively
pristine gas (primordial abundances), and will soon be detectable by
upcoming missions, such as JWST, through their supernova explosions (de
Souza et al. 2013, 2014). These first stars
(Pop-III) are thought to have been of high mass, although recent work suggests they fragmented possible into stars of several
tens of solar masses \citep[e.g.][]{2013MNRAS.433.1094S}. It is the {\sl uv}-radiation of these stars that is thought to couple the hydrogen spin temperature to that of the cold
gas. Emission from resulting X-ray binaries (XRB) could subsequently lead to X-ray heating. In addition, the first 
and second generation (Pop-II) stars are thought
to be responsible for the ultimate (re)ionization of nearly all neutral hydrogen during the \EoR. What the relative roles of 
Pop-III and Pop-II stars exactly are in these processes, i.e.\ W-F coupling, heating and ionization, is currently ill-understood and the processes
involved in the formation of Pop-III stars and whether they form the seeds for SMBHs is also not clear. At lower redshifts where the first
galaxies can be observed, however, it is clear that stars responsible for reionization must be in mini-haloes and in galaxies $\sim$500 times
fainter ($\sim$7\,mag.) than current observational limits \citep[e.g.][]{2006MNRAS.370..273D}. Hence enormous extrapolations are needed to make any inference from current
observations, let alone to redshifts even higher. Redshifted 21-cm emission observations clearly are important in setting limits on star formation
and the types of stars (or AGN) that partake in the early phases of the CD/EoR~eras. 

\item{\bf Mini-Quasars and AGN}

A second source of ionization and heating can be intermediate-mass black holes (IMBHs) and resulting mini-quasars due to accretion discs or 
Bondi accretion \citep[e.g.][]{2006NewAR..50..204D}. Whereas it is thought that mini-quasars are not the dominant source of ionizing photons, they can 
still play a secondary role and possibly be a source of harder X-ray photons that can more uniformly heat the IGM well above the CMB
temperature. These IMBHs are also thought to be the seed-BHs of present-day SMBHs and AGN, although the 
process of accretion appears to be super-Eddington to have them grow from 100 solar masses (expected from massive Pop-III stars) 
to $10^9$ solar masses in the AGN seen already at $z\sim 7$ \citep[][]{2011Natur.474..616M}.

\item{\bf X-ray Binaries}

Great uncertainty also remains whether X-ray binaries are sources of heating at high redshifts and whether heating takes
place on global scales. If heating is inefficient some IGM patches could remain very cold and below the CMB temperature 
causing the redshifted 21-cm brightness temperature to remain in absorption 
substantially impacting the level of strength of the total-intensity and fluctuation signals from neutral hydrogen. This could lead
to considerable effects even at very low redshifts during the \EoR\ \citep[e.g.][]{2008PhRvD..78j3511P}.

\item{\bf Dark-Matter Annihilation} 

Literately a "Dark Horse"\footnote{Benjamin Disraeli, in "The Young Duke" (1831)} is whether dark matter could be a source
of ionization or heating at very high redshifts during the Dark Ages, already influencing the evolution of the 
IGM well before the Cosmic Dawn \citep[e.g][]{2008PhRvD..78l3510T, 2010MNRAS.406.2605R}. 

\end{itemize}

\noi Having outlined a number of important eras as well as the physics and source responsible for the brightness temperature evolution
of neutral hydrogen (both spatially and in time), we now continue to discuss which observables the SKA can obtain
and use to constrain these processes and from it learn about the sources responsible. We also shortly mention possible synergies
with other (planned) facilities.

\section{Observables via the redshifted 21-cm brightness temperature}\label{sect:observables}

\noi Redshifted 21-cm emission from neutral hydrogen manifests itself in
multiple ways and various methods and techniques can be used to extract
information from it. In the following section we discuss a number of such
approaches, whereas in the subsequent section a short summary is given of
other observables (and facilities) that the 21-cm brightness temperature
could be cross-correlated with. We refer to
\citet[][]{2013ExA....36..235M}, as well as to Chang et al. (2015) and
Jelic et al. (2015) in this volume, for more details on the synergy between
the SKA and other observatories.

\subsection{Measures of the brightness temperature field}

\noi {\it All of the information one hopes to gain from the physics occurring during the Dark Ages, Cosmic Dawn and Epoch of Reionization
from neutral hydrogen is obtained via its redshifted 21-cm brightness temperature.} As shown in Section~\ref{sect:physics}, this temperature is set by a combination of 
many different physical processes which need to be disentangled. As of yet, it is not fully clear whether this can be done completely or not. 
In addition, it is likely that in the first experiments, also with the SKA1, only limited measurements of the brightness temperature
will be made. Below we summarize some of the standard measures of $T_{\rm b}$:

\begin{itemize}     

     \item {\bf  Total Intensity} Redshifted 21-cm emission of neutral hydrogen averaged inside "shells" of common redshift 
     (or observed frequency) are  close to 
     impossible to measure with interferometers since they measure
intensity differences (Subrahmanyan et al. 2015). 
     However, as seen in Section~\ref{sect:physics}, the 
     total expected brightness temperature is of order $\sim 27 x_{\rm HI} (1+\delta)$\,mK at $z\sim 10$ if $T_{\rm s} \gg T_{\rm CMB}$. When 
     averaging over large areas of the sky ($\gg 1$ degree) $\langle \delta \rangle \approx 0$ and assuming no reionization (i.e.\  $x_{\rm HI}=1$),
     one can determine the total intensity of the signal. This is "most easily" achieved with single receiver/dipole measurements
     and is discussed in much greater detail in Subrahmanyan et al.\ (2015). Measuring this signal as function of redshift allows
     one to learn about different physical processes such as \LyA\ coupling by the first stars, heating by (X-ray) sources,
     reionization, etc. At this point it is not clear whether the SKA will be capable of
     measuring this signal since it requires a very accurately calibrated instrument.
     
     \item {\bf  Variance or second moment} The second-order effect comes from variations in the brightness temperature due to 
     variations of $\delta$, $x_{\rm HI}$, $T_{\rm s}$ and $v_{||}$. These variations lead to a spatial variance in $T_{\rm b}$
     as function of observed frequency or redshift (see Mellema et al.\
2015 and Wyithe et al.\ 2015) 
     which can be measured in interferometric images, assuming that systematics and 
     thermal noise are under control and/or known (although thermal noise can be removed via cross-variance techniques over
     small time-frequency steps). By measuring the variance one can again learn how HI evolved over cosmic time, in 
     particular when reionization peaked (the contrast between HI and bubbles is very large and leads to a large variance). However,
     the shape of PDF($T_{\rm b}$) also teaches us about its moment (skewness and/or kurtosis) again revealing information
     about reionization \citep[e.g.][]{2009MNRAS.393.1449H}. In particular the skewness of the PDF might be a tell-tale signature of reionization that would be 
     hard to mimic by systematic effects (which often are strongly correlated between frequency and scale smoothly with frequency
     as well).
          
     \item {\bf (Anisotropic-)Power-spectra} One step further would be the analysis of $T_{\rm b}$ in the Fourier domain in terms of the 
     variance as function of spatial (or angular) scales, i.e.\ the power-spectrum. The power-spectrum depends on the spatially
     varying parameters in Equation~\ref{eqn:dt}, being $x_{\rm HI}$, $\delta$, $T_{\rm s}$ and $v_{||}$. All other parameters can be 
     assumed constant and only enter via the general FRW metric. The power-spectrum is directly related 
     to the underlying local physical processes and sources of reionization (governing $x_{\rm HI}$), heating (governing $T_{\rm s}$), 
     as well as to the evolution of the density field (governing $\delta$ and $v_{||}$). To first order one can Taylor expand equation~\ref{eqn:dt}
     in these parameters, Fourier transform the result, from which the power-spectrum can be formed. Since the parameters are not
     independent (e.g.\ an over-density can also lead to extra heating and ionization), the resulting 21-cm power-spectrum becomes a
     rather complex function of these parameters and their cross-correlations, and can become anisotropic. 
     These can either be determined via simplified schemes
     or via direct numerical simulations (which also contain simplified sub-grid physics).  Disentangling all effects from the power-spectrum
     as function of redshift is nontrivial and remains an ongoing and active research field. To first order, the power-spectrum is isotropic and can be
     spherically averaged (where the 
     frequency direction over some limited bandwidth is a proxy for comoving distance, hence inverse wavenumber), although in the case of 
     SKA (possibly even before) the sensitivity if sufficient to measure anisotropy in the power-spectrum which can be an exciting 
     measure of underlying physics \citep[e.g.][]{2015PhRvL.114j1303F}.
     Hence it is often better to express the power-spectrum in two dimensions, separating the wave vector in a component perpendicular to the line
     of sight and a component parallel to it, in particular to study peculiar velocities, the light-cone effect, as well as potentially identify still-remaining
     systematic errors in the data-processing.
     
     \item {\bf Higher-order statistics/n-point correlations:} Already touched up on above, the brightness temperature field after the 
     \DA\ will start to deviate from a Gaussian random field. This means that the power-spectrum or two-point correlation
     function will become an increasingly more incomplete description. The brightness temperature difference between two points 
     will deviate from a Gaussian distribution, and the three and high-point statistics will become non-zero \citep[][]{2007ApJ...662....1P, 2008PhRvD..77j3506C}. Studying these
     statistics can therefore provide additional information about physical processes, even if direct imaging might still be infeasible.
     
      \item {\bf Tomography/Images:} The ultimate goal of any HI observation is make direct images and determine the brightness temperature
      with high significance for each spatial-frequency cell (see Mellema
et al.\ 2015 and Wyithe et al.\ 2015 this volume). 
      This clearly contains the maximum amount of information obtainable from the 
      data-set, but interpretation is not always straightforward and the analysis of images has been a sorely neglected topic in the literature. 
      In fact, even from direct numerical simulations, results are often presented in limited statistical measures such as the power-spectrum.
      Developing new ideas on how to analyze brightness temperature images is therefore crucial before the SKA comes online since it
      will be the first instrument capable of imaging/tomography on scales of several arc minutes to degrees. Only in exceptional cases
      might one expect to do this pre-SKA.
      
      \item {\bf Bubbles and HI topology:} Whereas initially imaging of HI brightness temperature fluctuations at the level of several mK might
      not be feasible (e.g. in roll-out, etc), 
      the stark contrast of 20-30mK and size (ten(s) of Mpc) of ionized bubbles will probably allow them to be seen relatively
      easy on during the roll-out and later phases of SKA1 imaging. 
      Their distribution as function of size, shape, etc. will likely reveal swiftly how reionization proceeded
      (i.e. inside-out, or otherwise). It might also reveal the nature of the sources causing reionization, in particular if (mini-)quasar play a
      role. Although some of this information is encoded in the power-spectra (which forms from a combination of the non-ionized
      HI field times a mask where HI has disappeared) their precise sizes, shapes, etc. will provide a far richer source of information. In addition it
      might be possible to study the local velocity field structure around bubbles which might reveal additional information that the power-spectra 
      do not.  
      
      \item {\bf 21-cm absorption:} Whereas all the above measures of neutral hydrogen are based on its brightness temperature contrast with 
      the CMB, there could be radio sources (mJy or brighter) present at sufficiently high redshift that the 21-cm line could be see in absorption
      against such source (e.g.\ \citet[][]{2002ApJ...577...22C} and Ciardi
et al.\ 2015). A mJy source (e.g.\ AGN or GRB) of a few arcmin resolution at say 150MHz  would already be sufficient to measure the HI forest
      and from it (equivalent to the \LyA\ forest) measure the line-of-sight power-spectrum of neutral hydrogen. Although success depends on 
      the (still unknown) presence of radio sources at redshifts well beyond the end of reionization, it would provide a unique opportunity to measure
      $P_{21}(k)$ on $k$-modes far larger (i.e.\ scales far smaller) than accessible via imaging/tomography or power-spectra as discussed above. As such SKA1 and 2 could provide
      the first measures of mini-haloes during the \EoR\ something that even {\sl JWST} will find difficult to do. 
      
\end{itemize}

\subsection{Synergy with other surveys/Cross-correlations}

\noi Whereas SKA will characterize the brightness temperature in detail via
either power-spectra, direct imaging or 21-cm absorption possible to
redshifts of $z \sim 27$ during the Cosmic Dawn, and will be unique at
this, complementary observations can further enhance the capability to
extract information from this rich data-set. Among these are: (1) the
Planck (or future) CMB maps, (2) the Far-Infrared background or emitters,
(3) surveys of high-$z$ \LyA\ emitters, (4) high-$z$ Lyman-break
Galaxies/Dropouts, (5) high-$z$ SNae/GRB transients, (6) CO/CII emitters
and (7) DLAs. We refer to Chang et al.\ (2015) and Vibor et al.\ (2015) in
this volume for details. 

\section{Survey Design}\label{sect:surveydesign}

\noi In this section we outline a potential observational survey with SKA1 (and 2) to characterize the redshifted 21-cm brightness temperature of HI
between $z=27$ ($\nu$=50MHz) and z$\sim$6 (end the EoR), via its two main observable: power-spectra and tomography. 
We assume throughout that calibration and foreground subtraction
is done to a level accurate enough that it does not affect the resulting residual data-cubes (containing both the 21-cm signal
as well as noise). For more details on the latter we refer to \cite{2013ExA....36..235M}. Hence thermal noise and sample variance 
(in the case of power-spectra) are assumed to dominate the error budget in the power-spectra and thermal noise dominates 
the error-budget in the images.

\subsection{Scaling Relations for power-spectrum sensitivity}

\noi To understand the choices of survey specifications we need some guidance. This is most easily understood by assuming a 
uniform $uv$-coverage in an interferometric experiment. As shown in \citet[][]{2013ExA....36..235M} and based on \citet[][]{2006ApJ...653..815M}, the  power-spectrum error due
to thermal noise is then given by:
\begin{equation}
\Delta^2_{\rm Noise} = \left( \frac{2}{\pi} \right) k^{3/2}  \left[D_{\rm c}^2 \Delta D_{\rm c}  \Omega_{\rm FoV} / N_{\rm b} \right]^{1/2}
	\left(\frac{T_{\rm sys}}{\sqrt{B\, t_{\rm int}}} \right)^2  \left( \frac{A_{\rm core} A_{\rm eff}}{A_{\rm coll}^2}   \right)
\end{equation}  
The total error is obtained by adding the sample variance to this, which depends on the power-spectrum of the signal
itself and the number of independent samples per volume. Whereas this equation gives a reasonable idea of the level of thermal noise variance on 21-cm power-spectra, it can be tilted 
as function of $k=2 \pi/L$, depending on the $uv$-density as function of baseline length. In this equation, the co-moving distance
to the survey redshift is $D_{\rm c}$, the depth of the survey for a bandwidth $B$ is $\Delta D_{\rm c}$ and  
the field-of-view of the survey is $ \Omega_{\rm FoV}$. In case of multi-beaming, the number of independent beams is $N_{\rm b}$.
A system temperature of $T_{\rm sys}$ is assumed and an 
integration time $t_{\rm int}$. Each station/antenna is assumed to have an effective collecting area $A_{\rm eff}$, that 
corresponds to the field-of-view as $\Omega_{\rm FoV} \equiv \lambda^2/A_{\rm eff}$, where $\lambda$ is the wavelength at the 
center of the frequency band. The antennae are distributed over a core area $A_{\rm core}$ in such a way that the $uv$-density
is uniform and the total collecting area (the sum of all effective collecting areas of each receiver) is assumed to be $A_{\rm coll}$.
The equation has been shown to be correct via numerical integration, using the method outlined in \citet[][]{2006ApJ...653..815M}.
Regardless the assumption of a uniform $uv$-density (which affect the exponent in $k^{3/2}$ and the normalization), the scaling with the layout of the
array follows
\begin{equation}
\Delta^2_{\rm Noise} \propto \left( \frac{A_{\rm core} \sqrt{A_{\rm eff}}}{A_{\rm coll}^2}   \right)
	\propto \left( \frac{A_{\rm core}}{N_{\rm stat}^2 A_{\rm eff}^{3/2}}   \right)
	\propto \left( \frac{A_{\rm core}}{\sqrt{N_{\rm stat}} A_{\rm coll}^{3/2}}   \right)
\end{equation}  
where we assume $A_{\rm coll} = N_{\rm stat} A_{\rm eff}$ with $N_{\rm stat}$ stations/receivers. 
Given the above equation
and scaling relations, a number of conclusions can be drawn for the power-spectrum sensitivity at a given $k$-mode scale: 
(i) sensitivity scales most rapidly with collecting area, (ii) increasing field-of-view helps, but only with the square root of the 
field of view, and (iii) more compact arrays (still covering the $k$-modes of interest) are more sensitive. All current arrays
(i.e.\ MWA, LOFAR and PAPER) follow strategies that optimize (or maximize within budgetary limits) these three parameters.
SKA1 and 2 will both increase $A_{\rm coll}$ by 1-2 order of magnitude over all current arrays and minimize
$A_{\rm core}$ by maximizing the filling factor and placing much of the collecting area in a rather small core area. Finally,
based largely on computational limitations, the field of view of SKA is at least of order five degrees at 100 MHz, but during 
later phases (e.g.\ SKA2) could be increased through either multi-beaming or through reducing the beam-formed number
of receivers. Although in considering optimizing an array for CD/EoR science, one should not forget that 
the system also needs to be calibrated for instrumental and ionospheric errors, and foregrounds (compact and extended, 
in all Stokes parameters) need to be removed, which might require the use of long baselines. Any array, also SKA1 and later 2,
therefore preferably would consist of a compact core with "arms" extending to many tens of kilometer.

\subsection{Scaling relations for tomography/imaging}

\noi Similar to the power-spectrum sensitivity, one can also obtain a scaling relation for imaging or tomography. One readily
finds that the thermal noise inside a resolution element corresponding to a scale $k_\perp$ scales as:
\begin{eqnarray}
\Delta T & =  &\left( \frac{k_{\perp}}{2 \pi} \right)   \left[ D_{\rm c}^2  \times \Omega_{\rm FoV} \right]^{1/2}
	\left(\frac{T_{\rm sys}}{\sqrt{B\, t_{\rm int}}} \right)  \sqrt{\left( \frac{A_{\rm core} A_{\rm eff}}{A_{\rm coll}^2}   \right)}  \nonumber \\
	   & = & \left( \frac{k_{\perp}}{2 \pi} \right)   \left[ D_{\rm c}^2  \times \Omega_{\rm core} \right]^{1/2}
	\left(\frac{T_{\rm sys}}{\sqrt{B\, t_{\rm int}}} \right) f_{\rm fill}^{-1}, 
\end{eqnarray}
assuming a {\sl sinc}-function as model for the sky-components (i.e. a top-hat in uv-space).
In this case $A_{\rm eff}$ drops out of the equation, although we leave it in to retain a similar form to the equation for
the power-spectrum. We also see that on larger angular scales (i.e.\ smaller $k_\perp$), the noise decreases and that a more compact 
array helps to reduce the thermal noise per resolution element, by increasing the number density of visibilities per
$uv$-cell. The latter form of the equation expresses this through the filling factor ($f_{\rm fill}$) of the core area in terms of 
collecting area and the field-of-view in a resolution element of the core. Filtering longer baselines (or smoothing 
of the image) to a resolution corresponding to $k_\perp$ averages down the noise and lowers the brightness 
temperature error, hence $k_\perp$ does not necessarily correspond to the largest baseline afforded by the core area. The 
latter simply sets the $uv$-density. For example for the inner part of SKA1 the core of 400m diameter has 
$f_{\rm fill}\sim 1$ and on scales $k_\perp = 0.1$ cMpc$^{-1}$ ($\sim 20'$) one expect an error of $\sim 0.2$\,mK per MHz 
in 1000\,hrs of integration time at 150MHz, assuming $T_{\rm sys}=400$\,K, which is below the expected 
brightness temperature fluctuations on that scale. However, with increasing resolution the filling factor of the 
core decreases and $k_\perp$ increases, both leading to a rapidly increasing brightness temperature error ultimately
limiting the angular scale where imaging can still be done. For SKA1 and 2 this limits occurs on an angular scale of a few
to about ten arc minutes in a reasonable ($\sim$1000 hr) integration time.

\subsection{Current SKA1-LOW Baseline Design and its Relevant Design Specifications}\label{sect:BLD}

\noi In the following we assume the current SKA1 baselines design (BLD),
based on the station layout provided in Braun (2014).
Although the latter is not necessarily the ultimate layout, it represents a
"close-enough" description of the array that no major differences are
expected in the resulting sensitivity calculations, except if the array is
(i) substantially thinned out (i.e.\ reducing the total collecting area),
(ii) dramatic changes in antenna layout are introduced (i.e.\ changing the
filling factor) or (iii) substantially reducing the instantaneous field of
view (i.e.\ through beam forming larger numbers of receiver elements).  We
refer to other chapters in this volume for a full description of the
current BLD. Using these antenna positions (or radial distribution) we
calculate full 12-hr $uv$-tracks in the direction of the South Galactic
pole and average the $uv$-density within radial bins.  Similar densities
are obtained for tracks at lower elevations.  We assume a nominal 35m
beam-formed "station" as well, which leads to a frequency-dependent
field-of-view, being around 5 degrees FWHM at 100\,MHz.

\noi The fiducial layout that we assume in this review is listed below. We emphasize that this is not 
necessarily the most optimal observing strategy, although in some cases (e.g. freq. coverage) full use of the capabilities
of SKA1 and 2 are made. As we will discuss below, {\it multi-beaming can dramatically improve the return
of a single CD/EoR survey}. Also integration times can be increased\footnote{We note that largely only winter-nights can be 
used during which the Sun and Milky Way center are below the horizon {\sl and} the ionospheric effect are milder. Hence
for deep pointed integrations, accumulating a thousand hours of on-sky time might take multiple years as experience with 
current pathfinder/precursors of the SKA already shows. Leeway to offset collecting area against integration time 
might therefore not save costs since it can lead to extremely long project lifetimes.}. Finally we assume for the nominal
survey that phase-tracking of a single field is done, rather than drift-scans although we come back to that in this Section 
and Section~\ref{sect:mediumshallow}.

\begin{figure}
\hspace{-2.5cm}
\includegraphics[width=1.3\textwidth]{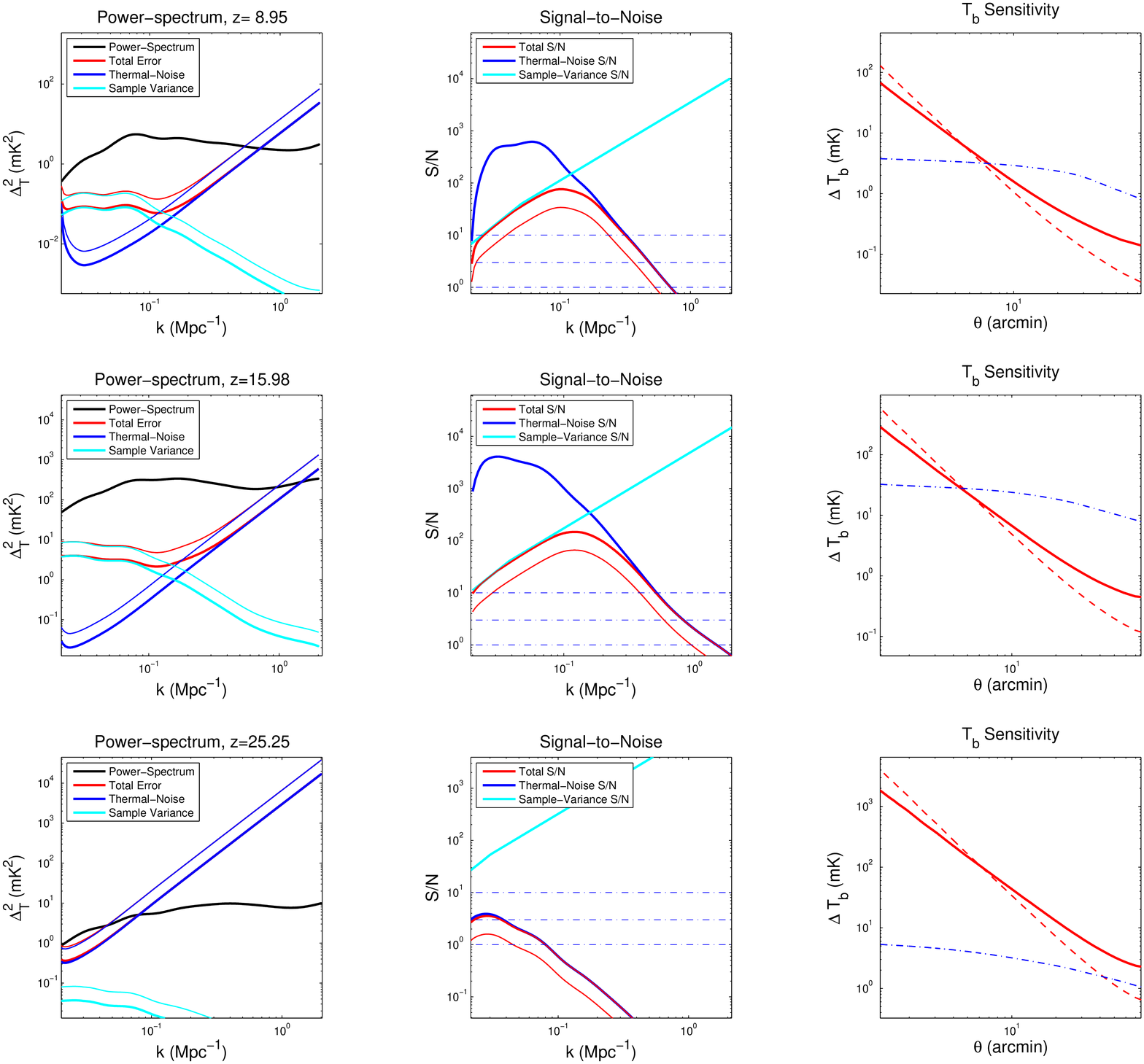}
\caption{{\bf Deep Survey:} Shown are the expected brightness temperature power-spectra from \citet[][]{2014MNRAS.439.3262M}
and the expected thermal-noise and cosmic-variance errors \citep[following][]{2006ApJ...653..815M} for the fiducial deep-survey parameters
outlined in the text for SKA1-LOW. The thick/thin curves are for 5 versus 1 beam. The middle-column panels (red curves) indicate the expected 
total S/N ratio in such a deep survey, reaching a peak S/N$\sim 70$, in good agreement with the results in \citet[][]{2014MNRAS.439.3262M}. 
The right column panels show the brightness temperature sensitivity as function of resolution, where the $k$-scale is transformed 
directly to angular scale at the corresponding redshift. The latter does not depend on the number of beams. The dashed
line is for a BW that matches the angular scale rather than being fixed to 10MHz. The blue dot-dashed line is the expected variance 
on that scale by integrating over the power-spectrum.}\label{fig:powerspectra1}
\end{figure}

\begin{enumerate}

\item {\bf Antenna layout:} We assume for SKA1 the layout as (generically) specified the BLD, assuming that for SKA2 the collecting 
area is increased by a factor of four. The latter is implemented here via a scaling of all baselines by a factor of two and
quadrupling the number of stations at each SKA1 station position. This ensures that the filling factor in the core remains of 
order unity and decreases in a scale-invariant manner to SKA1. Other choices are possible and will be investigated in the future. 

\item {\bf Frequency range:} Based on the current baseline design of SKA1, we assume a full instantaneous
frequency coverage of 50--350\,MHz (i.e.\ $z$=27-3). This range is thought to encompass the full \CD\ 
and \EoR\ \citep[see][]{2013ExA....36..235M}. We assume that during SKA1 the bandwidth can at least be split in two beams of 150-MHz,
increasing the survey speed by a factor two. We note again that more beams can reduce on-sky time significantly. 
In the future one might consider extending SKA1 to lower frequencies 
(below 50 MHz) during the upgrade/extension to SKA2, although this will not be considered in this chapter. 

\item {\bf Survey FoV}: The field of view of a single primary beam is assumed to be FWHM$\approx 5^\circ$ at 100\,MHz\footnote{We assume a D=35m beam-formed primary beam at 100\,MHz, although not necessarily physical stations of 35m.}. At a redshift of $z\sim 10$, the comoving
distance is $\sim 9700$\,cMpc, hence the comoving scales covered by the primary beam extend up to $\sim$1\,cGpc, about ten times
larger than the scale where the dark-matter power-spectrum peaks ($\sim 100$\,cMpc). For a bandwidth of 10 MHz (see below), the 
volume depth is $\sim 170$\,cMpc. This means that sample variance on scales of a few hundred cMpc is $\sim 10 \%$ and less on smaller
scales. On larger scales, sample variance rapidly increases and only a larger survey volume (i.e.\ increased field-of-view via multi-beaming
or drift-scanning) can further reduce this source of variance. One might consider increasing the FoV by beam forming a smaller number
of log-periodic elements. We discuss the effect of multi-beaming/increased FoV below.

\item {\bf Integration times}: In this review, we assume an integration time of 10, 100 and 1000\,hrs, for the different levels in a 
three-tiered survey with decreasing survey area, respectively. 
The reason for this choice is that SKA1 can reach 
levels of $\sim$1\,mK in 1000-hrs integration time (per 1-MHz bandwidth) over a wide range of angular scales. However, this integration 
should only be thought of as a reference, since the level of brightness temperature can vary between models. This
integration time is also often used in the literature as a reference, making comparison between models easier. The overall on-sky 
time of the assumed (three-tiered) survey, however, will be larger (see Section\,\ref{sect:survey}).

\item {\bf Frequency Bandwidth/Redshift Binning: } We assume a bandwidth of 10\,MHz, per redshift bin (not in total!), to maximize sensitivity but minimize light-cone effects, i.e.\
the impact of the evolution of the hydrogen brightness temperature with redshift mixed with the assumption that redshift can be regarded as a third 
spatial dimension of the data volume. The assumed bandwidth is a reasonable compromise. We note that bandwidth enters both 
in the survey depth (via $\Delta D_{\rm c}$) and in the sensitivity per visibility.

\item {\bf Multi-Beaming:} We assume a single primary beam ($N_{\rm b}=1$) for a 35m station in SKA1 for the full bandwidth (BW), possible two
when split in to two 150-MHz bands, increasing for SKA2 potentially to covering the full log-period dipole beam ($N_{\rm b} \gg 1$). The
latter would require an increase in correlator capacity beyond the currently envisioned SKA1 correlator. Below we will show the impact
of multi-beaming, in particular in dramatically reducing sample variance on large angular scales.

\item {\bf System Temperature:} We assume $T_{\rm sys} = 100 + 400\times (\nu/150{\,\rm MHz})^{-2.55}$\,K. Although this can change
from field to field and can increase/decrease towards/away from the Galactic plane, we assume it to hold to reasonable order and
is typical for assumptions in the literature.

\item {\bf Inner-Cut uv-plane:} Finally we assume a 30-$\lambda$ inner cut to the $uv$-plane (i.e. the sky-projected baselines in units of wavelength)
to reduce the impact for foreground 
leakage in to the power-spectrum. This starts limiting scales $k < 0.02$\,cMpc$^{-1}$ in the power-spectrum (See Fig.\ref{fig:powerspectra1}) 
corresponding to scales of about 2 degrees 
or $\sim 300$\,cMpc at 150MHz. Since these scales already well exceed the peak of the DM and brightness temperature power spectra, they provide some information about
cosmology, but far less information of the physics of CD/EoR. This cut is not needed if bright and large scale foregrounds can be controlled.

\end{enumerate}

\noi Whereas these assumptions, based on the BLD of SKA1 and 2 should not be regarded as definitive, they are a guide to typical 
observations specification, also currently used e.g.\ by MWA, LOFAR and PAPER, 
and thus serve as a reference. 

\section{A Three-tiered Cosmic Dawn and Epoch of Reionization Survey}\label{sect:survey}

\noi Having specified our assumed nominal observational settings, in this section, we introduce an outline for  a
three-tiered survey of redshifted 21-cm emission during the CD/EoR that can address the wide 
range of physics questions posed in Section~\ref{sect:physics}.

\subsection{Main Observation Targets:}

\noi It should be  stressed that observations of HI emission from the CD/EoR have thus far not been made (although upper limits
are currently being set) and any present-day expectations are based on relatively sparse observations in different wavelength regimes or through very different observables(see Section~\ref{sect:currentobs}). {\it Any CD/EoR-HI observational strategy should therefore cover as much of parameter space as possible (e.g. redshift, angular and frequency scales, field of view, etc), providing for surprises that the Universe 
might put up on us, obviously bounded by any reasonable estimate of what is possible.} Overall there are three general approaches to detecting
and characterizing neutral hydrogen: 

\begin{itemize}

\item via {\bf direct imaging (i.e.\ tomography)} of neutral hydrogen down to the 1$\sigma \approx  1$\,mK brightness temperature level on $>$5-arcmin scales at $z\approx 6$, rising to degree scales at $z$=28. 

\item via {\bf statistical (including power-spectrum) methods} to variance levels 1$\sigma \approx 0.01-1$\,mK$^2$ over the redshift regime $z$=6-28, respectively, at k$<$0.1 Mpc$^{-1}$
and to k$<$1 Mpc$^{-1}$ at $z<15$.

\item via {\bf 21-cm absorption line observations} against high-z radio sources, if present, with 1-5\,kHz spectral resolution (2-10 km/s at z=10) with S/N>5 on 1 mJy. 

\item Although SKA2 can attain SKA1's limits 4$\times$ faster or, in the same integration time go 16$\times$ deeper (for power-spectra), new discoveries with SKA1 are expected to guide SKA2 observations, leading to new observational targets. SKA2 can also image smaller ionized bubbles and sub-mK brightness-temperature fluctuations during the Cosmic Dawn, unfeasible with SKA1. An extension of the frequency range from SKA1 to SKA2 could also enable one to probe the late Dark Ages at $z\sim$30-40.

\end{itemize}

\noi The general target can be reached in a set of surveys and target fields in three tiers (depth versus area), which we will discuss in the 
following three sub-sections.

\subsection{Deep Survey} 

\noi The direct science goals of a Deep CD/EoR survey are to detect and characterize ionized structures and HI brightness temperature fluctuations on 5--300 arcmin scales (varying) over the EoR/CD redshift range $z=6-28$ to 1-mK brightness temperature level and from it derive the state, thermal history and chemistry of the IGM, study the first stars, black holes and galaxies and constrain cosmology, the physics of dark matter and gravity.
To reach this goal, deep 1000hr integrations are considered on 5 separate 20-square degree windows covering a total of 100 square-degree on-sky area, using the 50-200MHz frequency range with 0.1 MHz spectral resolution (for science), using the full SKA1-LOW array, using multi-beaming with $N_{\rm b}=2$ lowering the on-sky time from 5000 to 2500 hrs. This will not be feasible with any current (or funded) array with S/N$>$1 and is unique to SKA1-LOW. This survey lowers the (expected) thermal noise $\sim 10 \times$ over LOFAR and allows direct imaging with S/N$\gg$1. The Cosmic Dawn remains inaccessible to current/funded instruments, until this deep survey with SKA1-LOW. The sensitivities for power-spectra and imaging are shown in Fig.~ref{fig:powerspectra1}.

\subsection{Medium+Shallow Survey}\label{sect:mediumshallow}

\begin{figure}
\hspace{-2.5cm}
\includegraphics[width=1.3\textwidth]{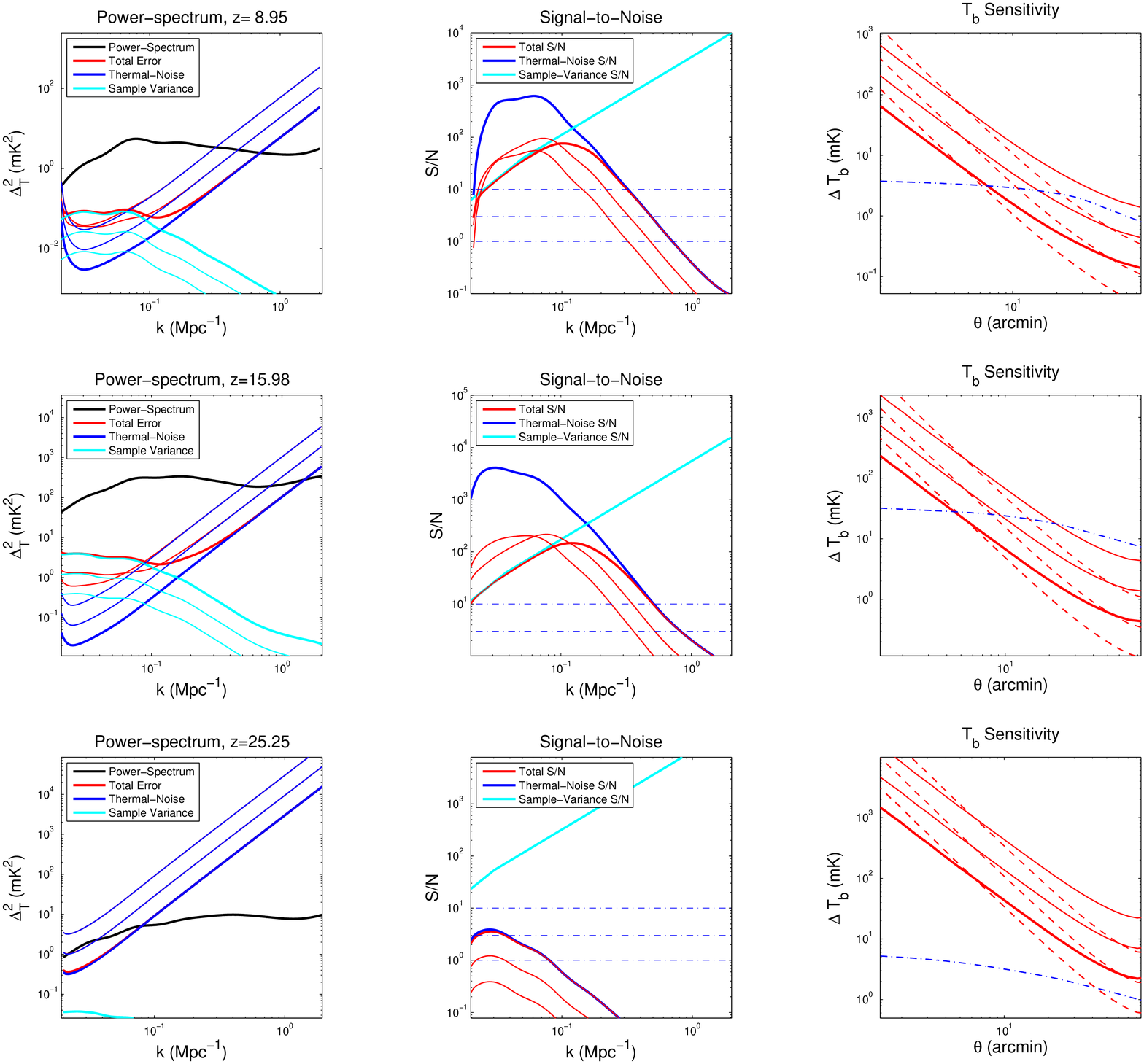}
\caption{{\bf Deep-Medium-Shallow Surveys:} Shown are the expected brightness temperature power-spectra from Mesinger et al.\ (2014)
and the expected thermal-noise and cosmic-variance errors (following McQuinn et al.\ 2006) for the fiducial survey parameters
outlined in Section~6 for SKA1. The middle panels (thick red curve) indicate the expected total S/N ratio in such a 
survey, reaching a peak S/N$\sim 70$, in good agreement with the results in Mesinger et al.\ (2014). From top to bottom, the red curves
for the deep, medium and shallow surveys are shown, clearly showing an increased gain in sensitivity at the larger $k$ modes in 
the deep survey due to low thermal noise, and an increased gain in sensitivity at the smaller $k$ modes in 
the medium+shallows surveys due to lower sample variance, leading to a relatively flat S/N curve over one dex in $k$ space. 
For completeness, the right column show the brightness temperature sensitivity as function of resolution, where $k$-scale is transformed 
directly to angular scale at the corresponding redshift. From bottom to top are the deep, medium and shallow surveys. The dashed
line is for a BW that matches the angular scale rather than being fixed to 10MHz. The blue dot-dashed line is the expected variance 
on that scale by integrating over the power-spectrum.}
\label{fig1}\label{fig:powerspectra2}
\end{figure}

\noi The direct science goals of a Medium+Shallow CD/EoR survey are to detect and characterize the 21-cm power-spectrum with a peak S/N$\sim$70 measured over $k=0.02-1.00$\,Mpc$^{-1}$ over the CD/EoR redshift range $z \sim 6-28$ and from it derive the state, thermal history and chemistry of the IGM, study the first stars, black holes and galaxies and constrain cosmology, the physics of dark matter and gravity.   
A medium-deep pointed survey of 50 times 100-hr integrations would cover 1,000 square degrees, and a preceding shallow pointed/drift-scan survey of 500 times 10-hr integrations would cover 10,000 square degrees, using the 50-200\,MHz frequency range with 0.1\,MHz spectral resolution (for science) and using the full SKA1-LOW array. Both survey can be carried out in 2500 hours each, assuming two beams of 150 MHz. Again 
multi-beaming could reduce the on-sky time substantially. A gain in power-spectrum sensitivity would be 1-2 orders of magnitude over 
current arrays (i.e.\ MWA, PAPER, LOFAR)  in the sample-variance limited S/N regime ($k < 0.2$\,Mpc$^{-1}$), but scales  $k=0.2-1$\,Mpc$^{-1}$ are likely to remain inaccessible until SKA1-LOW is build, because of severe thermal-noise limitations. Again the Cosmic Dawn remains inaccessible to current/funded instruments, until SKA1-LOW is realized. The sensitivities for power-spectra and imaging are shown in Fig.\ref{fig:powerspectra2}.
  
\subsection{Absorption at 21-cm}

\noi The science goal is to obtain 21-cm absorption line spectra from (rare) radio sources at $z>6$ to probe very small scales (i.e. $k \sim 1000$\,Mpc$^{-1}$) or virialized structures (mini-haloes) and from it derive the state, thermal history and chemistry of the IGM, study the first stars, black holes and galaxies and constrain cosmology, the physics of dark matter and gravity.    
This can be done through deep 1000-hrs integrations on selected sources
with a flux of $>1$\,mJy (possibly in the Deep Survey fields; see Ciardi et
al. 2015) with a spectral resolution of a few km/s (1-5\,khz at 150 MHz) over the 50-200 MHz frequency range ($z\sim 6-28$).
We note also here, that no observations of this kind have ever been done and are not expected to be feasible before SKA1-LOW is build, because of the same severe thermal noise limitations that currently limit direct imaging.

\subsection{Deep to Shallow Surveys: Multi-beaming and/or Drift-Scans}

\noi Optimal observing strategies will depend on many variables: ionospheric conditions, minimization of side-lobe leakage of 
strong sources, field in the zenith, low galactic brightness temperatures and polarization, optimal
$uv$-coverage, multi-beaming versus frequency coverage, inclusion of long baseline information, monitoring of the station
beams, RFI excision, drift-scans versus tracking, etc. 
For SKA1-LOW the field of view is typically 5-10 degrees and hence the sky drifts \citep[see][for details]{2014PASA...31...26T}
though a stationary beam (i.e. non-tracking) in about 0.5-1\,hr. Deep integrations of 1000+ hrs would therefore require 
1000-2000 nights for every single deep field, which is prohibitively long. Tracking the field over $N$\,hrs, would allow the same 
integration time to be accumulated in $N$ times fewer nights. Hence it appears that drift-scans would only be useful for larger 
scale shallow surveys, where the total integration per field can be small. On the other hand, because SKA1-LOW allows for two 
beams  in the baseline design, one could conceive of a combination of tracking-drift-scan, where the field drifts from one 
stationary beam to a second stationary beam and only every 0.5-1\,hr would one switch the phase center of the trailing beam
to a new phase-center. Since these ideas have not been worked out in detail  \citep[but see][]{2014PASA...31...26T}, the
precise observing strategy remains an open question and will depend on a combination of nights available, sensitivity, 
resolution in the image, calibration errors, etc.


\section{Role-out of SKA1 and expansion to SKA2.}\label{sect:rollout}

\noi A great advantage of radio interferometers is that it can observe even if incomplete in its number of receiver elements
and/or stations/dishes. To enable high-impact science even with an incomplete array it is important to think about rollout
from an incomplete SKA1 to a fully complete SKA2.

\subsection{Role-out from SKA1 to SKA2: 0.5, 1, 2 and 4 times SKA1}

\begin{figure}
\hspace{-2.5cm}
\includegraphics[width=1.3\textwidth]{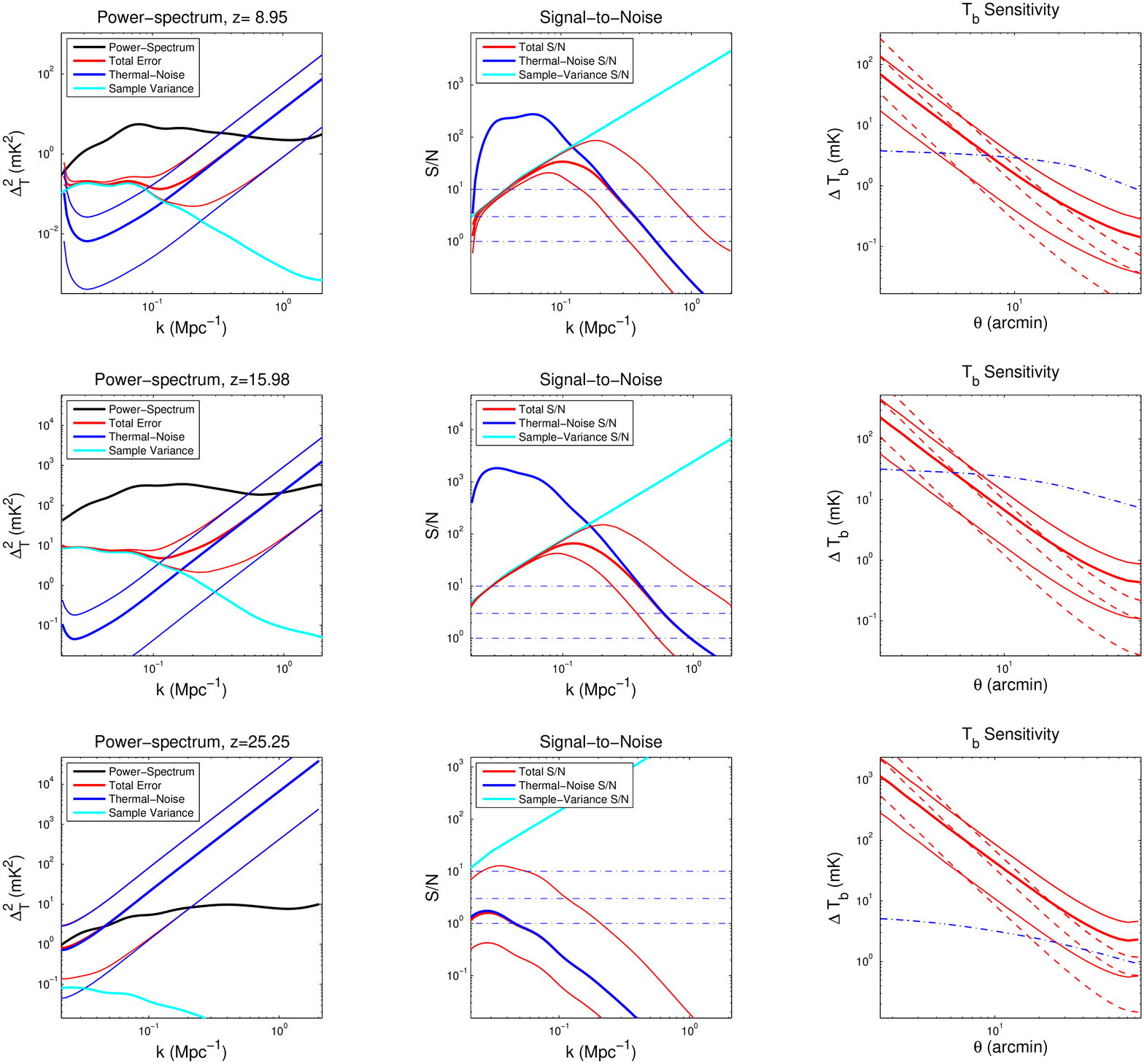}
\caption{{\bf Rollout from half to four times SKA1:} Idem to Fig.2 \& 3. The solid red lines from bottom to top (middle column) show the S/N
for a deep survey for $f_{\rm s}=0.5, 1.0, 4.0$. In the right column the brightness temperature (from bottom to top) is given for SKA2, SKA1
and half SKA1.}
\label{fig1}\label{fig:roleout}
\end{figure}

\noi In this section we shortly discuss three stages: (i) roll-out of half of SKA1, (ii) SKA1 and (iii) rollout to a full SKA2, being 
four times the collecting area of SKA1. We only compare these stagesca assuming a scaling in the number of stations
(i.e.\ collecting area) and no other specification (e.g. frequency coverage, multi-beaming, etc.). The ratio of the number of stations is 
assumed to be $f_{\rm s}$ where $f_{\rm s}=1$ for SKA1.  
Since neither half of SKA1 nor SKA2 have any specified baseline designs, we have to make an assumption. Based on the 
scaling relations in Sections~\ref{sect:surveydesign}, a high filling factor is critical to reach surface brightness levels of $\sim$1\,mK. The simplest
assumption is therefore to take the current SKA1 BLD and scale the number
of visibilities (i.e.\ the voltage/electric-field coherence between two receivers) by $f^2_{\rm s}$, but keeping the radial uv-density profile scale-invariant. The above scalings ensure that 
the filling factor at the center of the array remains the same during rollout which ensure maximum sensitivity on large
angular scales.
In Fig.\ref{fig:roleout} we show the impact of $f_{\rm s}=0.5, 1.0, 4.0$ on the power-spectra and brightness temperature
sensitivity for tomography. We conclude a few things in comparison to the SKA1 baseline design:

\begin{itemize}

\item For $k < 0.1$\,Mpc$^{-1}$ there is no power-spectrum sensitivity loss in a single beam ($\sim 20$ square degrees) 
because one is sample-variance limited. Above $k > 0.1$\,Mpc$^{-1}$, however thermal noise dominates the error budget 
and one quickly gains S/N. Hence SKA1 in roll-out could already carry out a shallow wide-field survey focussed 
on $k < 0.1$\,Mpc$^{-1}$.

\item During the EoR and especially during the CD, SKA1 to SKA2 start to reach $k > 1$\,Mpc$^{-1}$ with good S/N, which is the region 
where the power-spectrum can rise quickly due to physics on small angular scale (mini-haloes) and where small ionized 
bubble will exhibit features in the power-spectrum. We note an almost order of magnitude increase in sensitivity from SKA1 
to SKA2 at $k$-modes beyond $k \sim 0.2$\,Mpc$^{-1}$

\item Whereas half SKA1 will enable tomography on $\sim 5$ arcmin scale during the EoR, during the CD this ability is 
lost and only with some luck scales larger than half a degree might be imaged. SKA1 and in particular SKA2, however,
will enable tomography during the EoR, CD and late Dark Ages (the latter only for SKA2). As shown in Sections~\ref{sect:surveydesign}~\&~\ref{sect:survey},
sheer collecting area and its resulting instantaneous sensitivity is needed. This can only partly be compensated by 
integrating longer, but at the expensive of substantially increasing the project duration (possibly to a decade or more).

\end{itemize}

\noi A remarkable result is that for most of the redshift regime ($z \approx 9-15$) cosmic variance
limits the survey below $k\approx 0.2$Mpc$^{-1}$. We note that because both
the thermal and cosmic-variance errors scale identically with number of beams (see text), this transition scale 
is invariant under increasing the number of beams (i.e.\ multi-beaming), despite that the S/N increases
as $\sqrt{N_{\rm beam}}$.

\section{Summary, Conclusions and Highlights}\label{sect:end}

\noi {\sl SKA1-LOW and finally SKA2-LOW will be a transformational instrument in the study of the physics of the first stars, galaxies and black holes, 
dark-matter, the IGM, cosmology and possibly gravity during the first one billion years of the Universe, via statistical and
direct measurements (e.g.\ power-spectra and tomography) of the redshifted 21-cm line from neutral hydrogen. }

\noi SKA1 will enable brightness temperature levels of $\sim$1 mK to be reached in 1000 hrs of integration (BW=1\,MHz)  from 
$z \sim 6$ to $z\sim 28$ ($>$50\,MHz) over increasing angular scales ranging $\sim$5--300~arcminutes. SKA2 could potentially
enable even higher redshift observations, starting to probe the late Dark Ages, as well as probe a much larger area 
of the sky (via multi-beaming) to a much greater depth in less time, reducing the thermal noise variance that will still limit SKA1 on the 
small scales where the dominant source population (e.g. mini-haloes) have their largest impact on neutral hydrogen (both via ionization
and spin-temperature effects). The impact of SKA2 is expected in pushing tomography (i.e.\ direct imaging) to higher redshifts and to greater
S/N (e.g.\ to probe redshift space distortions, higher order statistics, specific regions in the cubes, etc) where the sensitivity
of SKA1 will still be insufficient. Although the \CD\ will be opened by SKA1, SKA2 will allow it to be studied in much great detail similar
to the impact of SKA1 on the study of neutral hydrogen during the \EoR.

\noi We have outlined a three-tiered survey (deep and shallow/medium-deep and 21-cm absorption line measurements) 
with SKA1 that allow one to reach the goals of measuring the HI brightness temperature to a level of around 1 mK for tomography 
(a genuinely unique capability of SKA1 and SKA2) over a range of a few arc minutes up to a degree (depending on redshift), reduce 
sample variance and reach power-spectrum measurements up to  one arcmin scales. Such a three tiered 
survey (assuming 21-cm absorption measurements can piggy-back on the deep survey) can be carried out in 
$3\times 2500$\,hrs\footnote{We note that collecting 2500 hrs of "good" deep pointed data might take up to $\sim$5 years since winter-nights 
where the MW and Sun are below the horizon and the ionosphere is mild might only allow up to 500 hrs on-sky time 
to be accumulated per calendar year. Multi-beaming  can significantly speed this up.} assuming that the full 300\,MHz BW of SKA1 
can be split in to two beams of 150\,MHz. Such a survey is expected to take of order five years, assuming an efficiency of 50\% of 
all night-time data (assuming 8-hr tracks) over such a period by selecting the best observing times and minimizing the impact of the 
ionosphere {\sl and} emission from the Milky Way. This is not a project that can be carried out in a few months of full-time observations.
Experience with the low-frequency SKA precursors and pathfinders confirms that good "observing conditions" are necessary 
for a successful CD/EoR projects. 

\noi Finally, a three-tiered survey with SKA1-LOW is expected to generate strong synergy with other major (future) facilities such 
as JWST, ALMA, Euclid, as well as with (CO/CII) intensity-mapping arrays and large ground-based IR facilities such as the ELT.
Many of the details can be found in the accompanying CD/EoR chapters in this volume.

\subsection{SKA1 and 2 Science High-lights}

\noi Some high-light science results, in random order, expected to be largely unique for SKA1 (and 2) are given below. For each
bullet point we refer to the first author of the related chapters in this volume:
\begin{itemize}
	\item Direct imaging of ionized regions and HI fluctuations on scales of arcminutes and larger during the \EoR\ and \CD\ ---
	{\sl Mellema et al. 2015; Wyithe et al. 2015}\ 
	\item Probing many aspects of cosmology/cosmography out to the highest possible redshifts (i.e.\ $z = 27$ in case of SKA1) --- {\sl Pritchard et al. 2015}\
	\item Enabling one to probe $ k \sim 1000$\,Mpc$^{-1}$ scales via 21-cm absorption, probing mini-halos even 
	out of reach of JWST --- {\sl Ciardi et al. 2015}\
	\item Direct study of the state and chemistry  of the IGM in the first billions years of the Universe -- {\sl  Ahn et al. 2015; Subrahmanyan et al. 2015}\ 
	\item Unique studies (i.e.\ CD) beyond the \EoR\ which will remain out of reach of most of not all other (planned) facilities
	in particular of the first (Pop-III stars and X-ray heating sources) -- {\sl Mesinger et al. 2015; Semelin et al. 2015}\
	\item Probing the impact of bulk-flows during the later \DA\ and the \CD\ allowing physics during the \DA\ and CMB to 
	be probed in a unique manner -- {\sl Maio et al. 2015}\ 
	\item Strong synergy with intensity mapping of the CO, CII, \LyA\ lines, possible molecular lines from primordial collapsing gas (e.g. H$_2$ and HD),  as well as the kS-Z effect and NIR/X-ray mission, as well as with many other (planned) facilities in space and on the ground --- {\sl Chang et al. 2015; Jelic et al. 2015}\
\end{itemize}

\noi We conclude to say that direct detection of neutral hydrogen at 
high redshift, beyond the statistical approaches of current experiments,  as well as probing the \CD\ will be defining abilities of the SKA (1 and 2). SKA can probe the universe back to 100\,Myr after the Big Bang, well beyond even the limitations of JWST\footnote{A single resolution element of SKA1
and 2 will be as large as the entire field of view of the JWST, placing "complementarity" in a different footlight. It is likely, if the 
two instruments have overlapping lifetimes that JWST will follow-up limited target regions selected by the SKA.}. This unique science can be
carried out in a three-tiered survey of 7500 hrs or total on-sky time.

\setlength{\bibsep}{0.0pt}
\bibliographystyle{truncapj}

\bibliography{reionization}

\begin{thebibliography}{61}
\expandafter\ifx\csname natexlab\endcsname\relax\def\natexlab#1{#1}\fi

\bibitem[{{Ahn}{et~al.}(2015){Ahn}}]{Ahnvol}{Ahn},
  K., et al., 2015, ``Probing the First Galaxies and Their Impact on the
Intergalactic Medium through 21-cm Observations of the Cosmic Dawn with the
SKA'', in proceedings of ``Advancing Astrophysics with the Square Kilometre Array", PoS(AASKA14)003

\bibitem[{{Ali} {et~al.}(2015){Ali}, {Parsons}, {Zheng}, {Pober}, {Liu},
  {Aguirre}, {Bradley}, {Bernardi}, {Carilli}, {Cheng}, {DeBoer}, {Dexter},
  {Grobbelaar}, {Horrell}, {Jacobs}, {Klima}, {MacMahon}, {Maree}, {Moore},
  {Razavi}, {Stefan}, {Walbrugh}, \& {Walker}}]{2015arXiv150206016A}
{Ali}, Z.~S., {Parsons}, A.~R., {Zheng}, H., {et~al.} 2015, ArXiv e-prints

\bibitem[{{Ali-Ha{\"i}moud} {et~al.}(2014){Ali-Ha{\"i}moud}, {Meerburg}, \&
  {Yuan}}]{2014PhRvD..89h3506A}
{Ali-Ha{\"i}moud}, Y., {Meerburg}, P.~D., \& {Yuan}, S. 2014, \prd, 89, 083506

\bibitem[{{Barkana} \& {Loeb}(2001)}]{2001PhR...349..125B}
{Barkana}, R., \& {Loeb}, A. 2001, \physrep, 349, 125

\bibitem[{{Becker} {et~al.}(2001){Becker}, {Fan}, {White}, {Strauss},
  {Narayanan}, {Lupton}, {Gunn}, {Annis}, {Bahcall}, {Brinkmann}, {Connolly},
  {Csabai}, {Czarapata}, {Doi}, {Heckman}, {Hennessy}, {Ivezi{\'c}}, {Knapp},
  {Lamb}, {McKay}, {Munn}, {Nash}, {Nichol}, {Pier}, {Richards}, {Schneider},
  {Stoughton}, {Szalay}, {Thakar}, \& {York}}]{2001AJ....122.2850B}
{Becker}, R.~H., {Fan}, X., {White}, R.~L., {et~al.} 2001, \aj, 122, 2850

\bibitem[{{Bolton} {et~al.}(2010){Bolton}, {Becker}, {Wyithe}, {Haehnelt}, \&
  {Sargent}}]{2010MNRAS.406..612B}
{Bolton}, J.~S., {Becker}, G.~D., {Wyithe}, J.~S.~B., {Haehnelt}, M.~G., \&
  {Sargent}, W.~L.~W. 2010, \mnras, 406, 612

\bibitem[{{Bouwens} {et~al.}(2012){Bouwens}, {Illingworth}, {Oesch}, {Trenti},
  {Labb{\'e}}, {Franx}, {Stiavelli}, {Carollo}, {van Dokkum}, \&
  {Magee}}]{2012ApJ...752L...5B}
{Bouwens}, R.~J., {Illingworth}, G.~D., {Oesch}, P.~A., {et~al.} 2012, \apjl,
  752, L5

\bibitem[{{Braun}(2014)}]{Braun14}
{Braun}, R., 2014, ``SKA1 Array Configurations'', Document Number
SKA-OFF.AG.CNF-SKO-TN-001 Revision 1, SKA Organisation

\bibitem[{{Carilli} {et~al.}(2002){Carilli}, {Gnedin}, \&
  {Owen}}]{2002ApJ...577...22C}
{Carilli}, C.~L., {Gnedin}, N.~Y., \& {Owen}, F. 2002, \apj, 577, 22

\bibitem[{{Cooray} {et~al.}(2008){Cooray}, {Li}, \&
  {Melchiorri}}]{2008PhRvD..77j3506C}
{Cooray}, A., {Li}, C., \& {Melchiorri}, A. 2008, \prd, 77, 103506

\bibitem[{{Chang}{et~al.}(2015){Chang}}]{Changvol}{Chang},
  T.C., et al., 2015, ``Synergy of CO/[CII]/Lya Line Intensity Mapping with
the SKA'', in proceedings of ``Advancing Astrophysics with the Square Kilometre Array", PoS(AASKA14)004

\bibitem[{{Chapman}{et~al.}(2015){Chapman}}]{Chapmanvol}{Chapman},
  E., et al., 2015, ``Cosmic Dawn and Epoch of Reionization Foreground
Removal with the SKA'', in proceedings of ``Advancing Astrophysics with the Square Kilometre Array", PoS(AASKA14)005

\bibitem[{{Ciardi}{et~al.}(2015){Ciardi}}]{Ciardivol}{Ciardi},
  B., et al., 2015, ``21-cm forest with the SKA'', in proceedings of ``Advancing Astrophysics with the Square Kilometre Array", PoS(AASKA14)006

\bibitem[{{Dav{\'e}} {et~al.}(2006){Dav{\'e}}, {Finlator}, \&
  {Oppenheimer}}]{2006MNRAS.370..273D}
{Dav{\'e}}, R., {Finlator}, K., \& {Oppenheimer}, B.~D. 2006, \mnras, 370, 273

\bibitem[de Souza et al.(2013)]{2013MNRAS.436.1555D} de Souza, R.~S., 
Ishida, E.~E.~O., Johnson, J.~L., Whalen, D.~J., 
\& Mesinger, A.\ 2013, \mnras, 436, 1555 

\bibitem[de Souza et al.(2014)]{2014MNRAS.442.1640D} de Souza, R.~S., 
Ishida, E.~E.~O., Whalen, D.~J., Johnson, J.~L., 
\& Ferrara, A.\ 2014, \mnras, 442, 1640 

\bibitem[{{Dijkstra}(2006)}]{2006NewAR..50..204D}
{Dijkstra}, M. 2006, \nar, 50, 204

\bibitem[{{Dillon} {et~al.}(2014){Dillon}, {Liu}, {Williams}, {Hewitt},
  {Tegmark}, {Morgan}, {Levine}, {Morales}, {Tingay}, {Bernardi}, {Bowman},
  {Briggs}, {Cappallo}, {Emrich}, {Mitchell}, {Oberoi}, {Prabu}, {Wayth}, \&
  {Webster}}]{2014PhRvD..89b3002D}
{Dillon}, J.~S., {Liu}, A., {Williams}, C.~L., {et~al.} 2014, \prd, 89, 023002

\bibitem[{{Ellis} {et~al.}(2013){Ellis}, {McLure}, {Dunlop}, {Robertson},
  {Ono}, {Schenker}, {Koekemoer}, {Bowler}, {Ouchi}, {Rogers}, {Curtis-Lake},
  {Schneider}, {Charlot}, {Stark}, {Furlanetto}, \&
  {Cirasuolo}}]{2013ApJ...763L...7E}
{Ellis}, R.~S., {McLure}, R.~J., {Dunlop}, J.~S., {et~al.} 2013, \apjl, 763, L7

\bibitem[{{Fan} {et~al.}(2002){Fan}, {Narayanan}, {Strauss}, {White}, {Becker},
  {Pentericci}, \& {Rix}}]{2002AJ....123.1247F}
{Fan}, X., {Narayanan}, V.~K., {Strauss}, M.~A., {et~al.} 2002, \aj, 123, 1247

\bibitem[{{Fan} {et~al.}(2006){Fan}, {Strauss}, {Becker}, {White}, {Gunn},
  {Knapp}, {Richards}, {Schneider}, {Brinkmann}, \&
  {Fukugita}}]{2006AJ....132..117F}
{Fan}, X., {Strauss}, M.~A., {Becker}, R.~H., {et~al.} 2006, \aj, 132, 117

\bibitem[{{Fernandez} \& {Zaroubi}(2013)}]{2013MNRAS.433.2047F}
{Fernandez}, E.~R., \& {Zaroubi}, S. 2013, \mnras, 433, 2047

\bibitem[{{Fialkov} {et~al.}(2015){Fialkov}, {Barkana}, \&
  {Cohen}}]{2015PhRvL.114j1303F}
{Fialkov}, A., {Barkana}, R., \& {Cohen}, A. 2015, Physical Review Letters,
  114, 101303

\bibitem[{{Fialkov} {et~al.}(2014{\natexlab{a}}){Fialkov}, {Barkana}, {Pinhas},
  \& {Visbal}}]{2014MNRAS.437L..36F}
{Fialkov}, A., {Barkana}, R., {Pinhas}, A., \& {Visbal}, E. 2014{\natexlab{a}},
  \mnras, 437, L36

\bibitem[{{Fialkov} {et~al.}(2014{\natexlab{b}}){Fialkov}, {Barkana}, \&
  {Visbal}}]{2014Natur.506..197F}
{Fialkov}, A., {Barkana}, R., \& {Visbal}, E. 2014{\natexlab{b}}, \nat, 506,
  197

\bibitem[{{Fialkov} \& {Loeb}(2013)}]{2013JCAP...11..066F}
{Fialkov}, A., \& {Loeb}, A. 2013, \jcap, 11, 66

\bibitem[{{Field}(1958)}]{1958PIRE...46..240F}
{Field}, G.~B. 1958, Proceedings of the IRE, 46, 240

\bibitem[{{Field}(1959)}]{1959ApJ...129..536F}
---. 1959, \apj, 129, 536

\bibitem[{{Furlanetto} \& {Briggs}(2004)}]{2004NewAR..48.1039F}
{Furlanetto}, S.~R., \& {Briggs}, F.~H. 2004, \nar, 48, 1039

\bibitem[{{Greenhill} {et~al.}(2014){Greenhill}, {Kocz}, {Barsdell}, {Clark},
  \& {LEDA Collaboration}}]{2014era..conf10301G}
{Greenhill}, L.~J., {Kocz}, J., {Barsdell}, B.~R., {Clark}, M.~A., \& {LEDA
  Collaboration}. 2014, in Exascale Radio Astronomy, 10301

\bibitem[{{Harker} {et~al.}(2009){Harker}, {Zaroubi}, {Thomas}, {Jeli{\'c}},
  {Labropoulos}, {Mellema}, {Iliev}, {Bernardi}, {Brentjens}, {de Bruyn},
  {Ciardi}, {Koopmans}, {Pandey}, {Pawlik}, {Schaye}, \&
  {Yatawatta}}]{2009MNRAS.393.1449H}
{Harker}, G.~J.~A., {Zaroubi}, S., {Thomas}, R.~M., {et~al.} 2009, \mnras, 393,
  1449

\bibitem[{{Helgason} {et~al.}(2014){Helgason}, {Cappelluti}, {Hasinger},
  {Kashlinsky}, \& {Ricotti}}]{2014ApJ...785...38H}
{Helgason}, K., {Cappelluti}, N., {Hasinger}, G., {Kashlinsky}, A., \&
  {Ricotti}, M. 2014, \apj, 785, 38

\bibitem[{{Iliev}{et~al.}(2015){Iliev}}]{Ilievvol}{Iliev},
  I., et al., 2015, ``Epoch of Reionization modelling and simulations for
SKA'', in proceedings of ``Advancing Astrophysics with the Square Kilometre Array", PoS(AASKA14)007

\bibitem[{{Jelic}{et~al.}(2015){Jelic}}]{Jelicvol}{Jelic},
  V., et al., 2015, ``SKA - EoR correlations and cross-correlations: kSZ,
radio galaxies, and NIR background'', in proceedings of ``Advancing Astrophysics with the Square Kilometre Array", PoS(AASKA14)008

\bibitem[{{Lewis} \& {Challinor}(2007)}]{2007PhRvD..76h3005L}
{Lewis}, A., \& {Challinor}, A. 2007, \prd, 76, 083005

\bibitem[{{Loeb} \& {Furlanetto}(2013)}]{2013fgu..book.....L}
{Loeb}, A., \& {Furlanetto}, S.~R. 2013, {The First Galaxies in the Universe}

\bibitem[{{Lorenzoni} {et~al.}(2011){Lorenzoni}, {Bunker}, {Wilkins},
  {Stanway}, {Jarvis}, \& {Caruana}}]{2011MNRAS.414.1455L}
{Lorenzoni}, S., {Bunker}, A.~J., {Wilkins}, S.~M., {et~al.} 2011, \mnras, 414,
  1455

\bibitem[{{Madau} {et~al.}(1997){Madau}, {Meiksin}, \&
  {Rees}}]{1997ApJ...475..429M}
{Madau}, P., {Meiksin}, A., \& {Rees}, M.~J. 1997, \apj, 475, 429

\bibitem[{{Maio}{et~al.}(2015){Maio}}]{Maiovol}{Maio},
  U., et al., 2015, ``Bulk Flows and End of the Dark Ages with the SKA'', in proceedings of ``Advancing Astrophysics with the Square Kilometre Array", PoS(AASKA14)009

\bibitem[{{Matthee} {et~al.}(2014){Matthee}, {Sobral}, {Swinbank}, {Smail},
  {Best}, {Kim}, {Franx}, {Milvang-Jensen}, \& {Fynbo}}]{2014MNRAS.440.2375M}
{Matthee}, J.~J.~A., {Sobral}, D., {Swinbank}, A.~M., {et~al.} 2014, \mnras,
  440, 2375

\bibitem[{{McQuinn} {et~al.}(2006){McQuinn}, {Zahn}, {Zaldarriaga},
  {Hernquist}, \& {Furlanetto}}]{2006ApJ...653..815M}
{McQuinn}, M., {Zahn}, O., {Zaldarriaga}, M., {Hernquist}, L., \& {Furlanetto},
  S.~R. 2006, \apj, 653, 815

\bibitem[{{Mellema} {et~al.}(2013){Mellema}, {Koopmans}, {Abdalla}, {Bernardi},
  {Ciardi}, {Daiboo}, {de Bruyn}, {Datta}, {Falcke}, {Ferrara}, {Iliev},
  {Iocco}, {Jeli{\'c}}, {Jensen}, {Joseph}, {Labroupoulos}, {Meiksin},
  {Mesinger}, {Offringa}, {Pandey}, {Pritchard}, {Santos}, {Schwarz},
  {Semelin}, {Vedantham}, {Yatawatta}, \& {Zaroubi}}]{2013ExA....36..235M}
{Mellema}, G., {Koopmans}, L.~V.~E., {Abdalla}, F.~A., {et~al.} 2013,
  Experimental Astronomy, 36, 235

\bibitem[{{Mellema}{et~al.}(2015){Mellema}}]{Mellemavol}{Mellema},
  G., et al., 2015, ``HI tomographic imaging of the Cosmic Dawn and Epoch
of Reionization with SKA'', in proceedings of ``Advancing Astrophysics with the Square Kilometre Array", PoS(AASKA14)010

\bibitem[{{Mesinger}(2010)}]{2010MNRAS.407.1328M}
{Mesinger}, A. 2010, \mnras, 407, 1328

\bibitem[{{Mesinger} {et~al.}(2014){Mesinger}, {Ewall-Wice}, \&
  {Hewitt}}]{2014MNRAS.439.3262M}
{Mesinger}, A., {Ewall-Wice}, A., \& {Hewitt}, J. 2014, \mnras, 439, 3262

\bibitem[{{Mesinger}{et~al.}(2015){Mesinger}}]{Mesingervol}{Mesinger},
  A., et al., 2015, ``Constraining the Astrophysics of the Cosmic Dawn and
the Epoch of Reionization with the SKA'', in proceedings of ``Advancing Astrophysics with the Square Kilometre Array", PoS(AASKA14)011

\bibitem[{{Morales} \& {Wyithe}(2010)}]{2010ARA&A..48..127M}
{Morales}, M.~F., \& {Wyithe}, J.~S.~B. 2010, \araa, 48, 127

\bibitem[{{Mortlock} {et~al.}(2011){Mortlock}, {Warren}, {Venemans}, {Patel},
  {Hewett}, {McMahon}, {Simpson}, {Theuns}, {Gonz{\'a}les-Solares}, {Adamson},
  {Dye}, {Hambly}, {Hirst}, {Irwin}, {Kuiper}, {Lawrence}, \&
  {R{\"o}ttgering}}]{2011Natur.474..616M}
{Mortlock}, D.~J., {Warren}, S.~J., {Venemans}, B.~P., {et~al.} 2011, \nat,
  474, 616

\bibitem[{{Novaco} \& {Brown}(1978)}]{1978ApJ...221..114N}
{Novaco}, J.~C., \& {Brown}, L.~W. 1978, \apj, 221, 114

\bibitem[{{Paciga} {et~al.}(2011){Paciga}, {Chang}, {Gupta}, {Nityanada},
  {Odegova}, {Pen}, {Peterson}, {Roy}, \& {Sigurdson}}]{2011MNRAS.413.1174P}
{Paciga}, G., {Chang}, T.-C., {Gupta}, Y., {et~al.} 2011, \mnras, 413, 1174

\bibitem[Pacucci et al.(2014)]{2014MNRAS.443..678P} Pacucci, F., Mesinger, 
A., Mineo, S., \& Ferrara, A.\ 2014, \mnras, 443, 678 

\bibitem[{{Patra} {et~al.}(2013){Patra}, {Subrahmanyan}, {Raghunathan}, \&
  {Udaya Shankar}}]{2013ExA....36..319P}
{Patra}, N., {Subrahmanyan}, R., {Raghunathan}, A., \& {Udaya Shankar}, N.
  2013, Experimental Astronomy, 36, 319

\bibitem[{{Patra} {et~al.}(2015){Patra}, {Subrahmanyan}, {Sethi}, {Udaya
  Shankar}, \& {Raghunathan}}]{2015ApJ...801..138P}
{Patra}, N., {Subrahmanyan}, R., {Sethi}, S., {Udaya Shankar}, N., \&
  {Raghunathan}, A. 2015, \apj, 801, 138

\bibitem[{{Pillepich} {et~al.}(2007){Pillepich}, {Porciani}, \&
  {Matarrese}}]{2007ApJ...662....1P}
{Pillepich}, A., {Porciani}, C., \& {Matarrese}, S. 2007, \apj, 662, 1

\bibitem[{{Pober} {et~al.}(2015){Pober}, {Ali}, {Parsons}, {McQuinn},
  {Aguirre}, {Bernardi}, {Bradley}, {Carilli}, {Cheng}, {DeBoer}, {Dexter},
  {Furlanetto}, {Grobbelaar}, {Horrell}, {Jacobs}, {Klima}, {Kohn}, {Liu},
  {MacMahon}, {Maree}, {Mesinger}, {Moore}, {Razavi-Ghods}, {Stefan},
  {Walbrugh}, {Walker}, \& {Zheng}}]{2015arXiv150300045P}
{Pober}, J.~C., {Ali}, Z.~S., {Parsons}, A.~R., {et~al.} 2015, ArXiv e-prints

\bibitem[{{Pritchard} \& {Loeb}(2008)}]{2008PhRvD..78j3511P}
{Pritchard}, J.~R., \& {Loeb}, A. 2008, \prd, 78, 103511

\bibitem[{{Pritchard}{et~al.}(2015){Pritchard}}]{Pritchardvol}{Pritchard},
  J., et al., 2015, ``Cosmology from EoR/Cosmic Dawn with the SKA'', in proceedings of ``Advancing Astrophysics with the Square Kilometre Array", PoS(AASKA14)012

\bibitem[{{Ripamonti} {et~al.}(2010){Ripamonti}, {Iocco}, {Ferrara},
  {Schneider}, {Bressan}, \& {Marigo}}]{2010MNRAS.406.2605R}
{Ripamonti}, E., {Iocco}, F., {Ferrara}, A., {et~al.} 2010, \mnras, 406, 2605

\bibitem[{{Robertson} {et~al.}(2010){Robertson}, {Ellis}, {Dunlop}, {McLure},
  \& {Stark}}]{2010Natur.468...49R}
{Robertson}, B.~E., {Ellis}, R.~S., {Dunlop}, J.~S., {McLure}, R.~J., \&
  {Stark}, D.~P. 2010, \nat, 468, 49

\bibitem[{{Robertson} {et~al.}(2015){Robertson}, {Ellis}, {Furlanetto}, \&
  {Dunlop}}]{2015ApJ...802L..19R}
{Robertson}, B.~E., {Ellis}, R.~S., {Furlanetto}, S.~R., \& {Dunlop}, J.~S.
  2015, \apjl, 802, L19

\bibitem[{{Ruiz-Velasco} {et~al.}(2007){Ruiz-Velasco}, {Swan}, {Troja},
  {Malesani}, {Fynbo}, {Starling}, {Xu}, {Aharonian}, {Akerlof}, {Andersen},
  {Ashley}, {Barthelmy}, {Bersier}, {Castro Cer{\'o}n}, {Castro-Tirado},
  {Gehrels}, {G{\"o}{\v g}{\"u}{\c s}}, {Gorosabel}, {Guidorzi}, {G{\"u}ver},
  {Hjorth}, {Horns}, {Huang}, {Jakobsson}, {Jensen}, {K{\i}z{\i}lo{\v g}lu},
  {Kouveliotou}, {Krimm}, {Ledoux}, {Levan}, {Marsh}, {McKay}, {Melandri},
  {Milvang-Jensen}, {Mundell}, {O'Brien}, {{\"O}zel}, {Phillips}, {Quimby},
  {Rowell}, {Rujopakarn}, {Rykoff}, {Schaefer}, {Sollerman}, {Tanvir},
  {Th{\"o}ne}, {Urata}, {Vestrand}, {Vreeswijk}, {Watson}, {Wheeler}, {Wijers},
  {Wren}, {Yost}, {Yuan}, {Zhai}, \& {Zheng}}]{2007ApJ...669....1R}
{Ruiz-Velasco}, A.~E., {Swan}, H., {Troja}, E., {et~al.} 2007, \apj, 669, 1

\bibitem[{{Semelin}{et~al.}(2015){Semelin}}]{Semelinvol}{Semelin},
  B., et al., 2015, ``The physics of Reionization: processes relevant for
SKA observations'', in proceedings of ``Advancing Astrophysics with the Square Kilometre Array", PoS(AASKA14)013

\bibitem[{{Simcoe} {et~al.}(2012){Simcoe}, {Sullivan}, {Cooksey}, {Kao},
  {Matejek}, \& {Burgasser}}]{2012Natur.492...79S}
{Simcoe}, R.~A., {Sullivan}, P.~W., {Cooksey}, K.~L., {et~al.} 2012, \nat, 492,
  79

\bibitem[{{Smoot} {et~al.}(1992){Smoot}, {Bennett}, {Kogut}, {Wright}, {Aymon},
  {Boggess}, {Cheng}, {de Amici}, {Gulkis}, {Hauser}, {Hinshaw}, {Jackson},
  {Janssen}, {Kaita}, {Kelsall}, {Keegstra}, {Lineweaver}, {Loewenstein},
  {Lubin}, {Mather}, {Meyer}, {Moseley}, {Murdock}, {Rokke}, {Silverberg},
  {Tenorio}, {Weiss}, \& {Wilkinson}}]{1992ApJ...396L...1S}
{Smoot}, G.~F., {Bennett}, C.~L., {Kogut}, A., {et~al.} 1992, \apjl, 396, L1

\bibitem[{{Sobacchi} \& {Mesinger}(2014)}]{2014MNRAS.440.1662S}
{Sobacchi}, E., \& {Mesinger}, A. 2014, \mnras, 440, 1662

\bibitem[{{Sparre} {et~al.}(2014){Sparre}, {Hartoog}, {Kr{\"u}hler}, {Fynbo},
  {Watson}, {Wiersema}, {D'Elia}, {Zafar}, {Afonso}, {Covino}, {de Ugarte
  Postigo}, {Flores}, {Goldoni}, {Greiner}, {Hjorth}, {Jakobsson}, {Kaper},
  {Klose}, {Levan}, {Malesani}, {Milvang-Jensen}, {Nardini}, {Piranomonte},
  {Sollerman}, {S{\'a}nchez-Ram{\'{\i}}rez}, {Schulze}, {Tanvir}, {Vergani}, \&
  {Wijers}}]{2014ApJ...785..150S}
{Sparre}, M., {Hartoog}, O.~E., {Kr{\"u}hler}, T., {et~al.} 2014, \apj, 785,
  150

\bibitem[{{Stacy} \& {Bromm}(2013)}]{2013MNRAS.433.1094S}
{Stacy}, A., \& {Bromm}, V. 2013, \mnras, 433, 1094

\bibitem[{{Subrahmanyan}{et~al.}(2015){Subrahmanyan}}]{Subrahmanyanvol}{Subramanyan},
  R., et al., 2015, ``All-sky signals from recombination to reionization
with the SKA'', in proceedings of ``Advancing Astrophysics with the Square Kilometre Array", PoS(AASKA14)014

\bibitem[{{Taoso} {et~al.}(2008){Taoso}, {Bertone}, {Meynet}, \&
  {Ekstr{\"o}m}}]{2008PhRvD..78l3510T}
{Taoso}, M., {Bertone}, G., {Meynet}, G., \& {Ekstr{\"o}m}, S. 2008, \prd, 78,
  123510

\bibitem[{{Theuns} {et~al.}(2002){Theuns}, {Zaroubi}, {Kim}, {Tzanavaris}, \&
  {Carswell}}]{2002MNRAS.332..367T}
{Theuns}, T., {Zaroubi}, S., {Kim}, T.-S., {Tzanavaris}, P., \& {Carswell},
  R.~F. 2002, \mnras, 332, 367

\bibitem[{{Trott}(2014)}]{2014PASA...31...26T}
{Trott}, C.~M. 2014, \pasa, 31, 26

\bibitem[{{Tseliakhovich} \& {Hirata}(2010)}]{2010PhRvD..82h3520T}
{Tseliakhovich}, D., \& {Hirata}, C. 2010, \prd, 82, 083520

\bibitem[{{Vald{\'e}s} {et~al.}(2013){Vald{\'e}s}, {Evoli}, {Mesinger},
  {Ferrara}, \& {Yoshida}}]{2013MNRAS.429.1705V}
{Vald{\'e}s}, M., {Evoli}, C., {Mesinger}, A., {Ferrara}, A., \& {Yoshida}, N.
  2013, \mnras, 429, 1705

\bibitem[{{van Haarlem} {et~al.}(2013){van Haarlem}, {Wise}, {Gunst}, {Heald},
  {McKean}, {Hessels}, {de Bruyn}, {Nijboer}, {Swinbank}, {Fallows},
  {Brentjens}, {Nelles}, {Beck}, {Falcke}, {Fender}, {H{\"o}randel},
  {Koopmans}, {Mann}, {Miley}, {R{\"o}ttgering}, {Stappers}, {Wijers},
  {Zaroubi}, {van den Akker}, {Alexov}, {Anderson}, {Anderson}, {van Ardenne},
  {Arts}, {Asgekar}, {Avruch}, {Batejat}, {B{\"a}hren}, {Bell}, {Bell}, {van
  Bemmel}, {Bennema}, {Bentum}, {Bernardi}, {Best}, {B{\^i}rzan}, {Bonafede},
  {Boonstra}, {Braun}, {Bregman}, {Breitling}, {van de Brink}, {Broderick},
  {Broekema}, {Brouw}, {Br{\"u}ggen}, {Butcher}, {van Cappellen}, {Ciardi},
  {Coenen}, {Conway}, {Coolen}, {Corstanje}, {Damstra}, {Davies}, {Deller},
  {Dettmar}, {van Diepen}, {Dijkstra}, {Donker}, {Doorduin}, {Dromer}, {Drost},
  {van Duin}, {Eisl{\"o}ffel}, {van Enst}, {Ferrari}, {Frieswijk}, {Gankema},
  {Garrett}, {de Gasperin}, {Gerbers}, {de Geus}, {Grie{\ss}meier}, {Grit},
  {Gruppen}, {Hamaker}, {Hassall}, {Hoeft}, {Holties}, {Horneffer}, {van der
  Horst}, {van Houwelingen}, {Huijgen}, {Iacobelli}, {Intema}, {Jackson},
  {Jelic}, {de Jong}, {Juette}, {Kant}, {Karastergiou}, {Koers}, {Kollen},
  {Kondratiev}, {Kooistra}, {Koopman}, {Koster}, {Kuniyoshi}, {Kramer},
  {Kuper}, {Lambropoulos}, {Law}, {van Leeuwen}, {Lemaitre}, {Loose}, {Maat},
  {Macario}, {Markoff}, {Masters}, {McFadden}, {McKay-Bukowski}, {Meijering},
  {Meulman}, {Mevius}, {Middelberg}, {Millenaar}, {Miller-Jones}, {Mohan},
  {Mol}, {Morawietz}, {Morganti}, {Mulcahy}, {Mulder}, {Munk}, {Nieuwenhuis},
  {van Nieuwpoort}, {Noordam}, {Norden}, {Noutsos}, {Offringa}, {Olofsson},
  {Omar}, {Orr{\'u}}, {Overeem}, {Paas}, {Pandey-Pommier}, {Pandey}, {Pizzo},
  {Polatidis}, {Rafferty}, {Rawlings}, {Reich}, {de Reijer}, {Reitsma},
  {Renting}, {Riemers}, {Rol}, {Romein}, {Roosjen}, {Ruiter}, {Scaife}, {van
  der Schaaf}, {Scheers}, {Schellart}, {Schoenmakers}, {Schoonderbeek},
  {Serylak}, {Shulevski}, {Sluman}, {Smirnov}, {Sobey}, {Spreeuw}, {Steinmetz},
  {Sterks}, {Stiepel}, {Stuurwold}, {Tagger}, {Tang}, {Tasse}, {Thomas},
  {Thoudam}, {Toribio}, {van der Tol}, {Usov}, {van Veelen}, {van der Veen},
  {ter Veen}, {Verbiest}, {Vermeulen}, {Vermaas}, {Vocks}, {Vogt}, {de Vos},
  {van der Wal}, {van Weeren}, {Weggemans}, {Weltevrede}, {White}, {Wijnholds},
  {Wilhelmsson}, {Wucknitz}, {Yatawatta}, {Zarka}, {Zensus}, \& {van
  Zwieten}}]{2013A&A...556A...2V}
{van Haarlem}, M.~P., {Wise}, M.~W., {Gunst}, A.~W., {et~al.} 2013, \aap, 556,
  A2

\bibitem[{{Wouthuysen}(1952)}]{1952AJ.....57R..31W}
{Wouthuysen}, S.~A. 1952, \aj, 57, 31

\bibitem[{{Wyithe}{et~al.}(2015){Wyithe}}]{Wyithevol}{Wyithe},
  S., et al., 2015, ``Imaging HII Regions from Galaxies and Quasars During
Reionisation with SKA'', in proceedings of ``Advancing Astrophysics with the Square Kilometre Array", PoS(AASKA14)015

\bibitem[{{Yatawatta} {et~al.}(2013){Yatawatta}, {de Bruyn}, {Brentjens},
  {Labropoulos}, {Pandey}, {Kazemi}, {Zaroubi}, {Koopmans}, {Offringa},
  {Jeli{\'c}}, {Martinez Rubi}, {Veligatla}, {Wijnholds}, {Brouw}, {Bernardi},
  {Ciardi}, {Daiboo}, {Harker}, {Mellema}, {Schaye}, {Thomas}, {Vedantham},
  {Chapman}, {Abdalla}, {Alexov}, {Anderson}, {Avruch}, {Batejat}, {Bell},
  {Bell}, {Bentum}, {Best}, {Bonafede}, {Bregman}, {Breitling}, {van de Brink},
  {Broderick}, {Br{\"u}ggen}, {Conway}, {de Gasperin}, {de Geus}, {Duscha},
  {Falcke}, {Fallows}, {Ferrari}, {Frieswijk}, {Garrett}, {Griessmeier},
  {Gunst}, {Hassall}, {Hessels}, {Hoeft}, {Iacobelli}, {Juette},
  {Karastergiou}, {Kondratiev}, {Kramer}, {Kuniyoshi}, {Kuper}, {van Leeuwen},
  {Maat}, {Mann}, {McKean}, {Mevius}, {Mol}, {Munk}, {Nijboer}, {Noordam},
  {Norden}, {Orru}, {Paas}, {Pandey-Pommier}, {Pizzo}, {Polatidis}, {Reich},
  {R{\"o}ttgering}, {Sluman}, {Smirnov}, {Stappers}, {Steinmetz}, {Tagger},
  {Tang}, {Tasse}, {ter Veen}, {Vermeulen}, {van Weeren}, {Wise}, {Wucknitz},
  \& {Zarka}}]{2013A&A...550A.136Y}
{Yatawatta}, S., {de Bruyn}, A.~G., {Brentjens}, M.~A., {et~al.} 2013, \aap,
  550, A136

\bibitem[{{Zemcov} {et~al.}(2014){Zemcov}, {Smidt}, {Arai}, {Bock}, {Cooray},
  {Gong}, {Kim}, {Korngut}, {Lam}, {Lee}, {Matsumoto}, {Matsuura}, {Nam},
  {Roudier}, {Tsumura}, \& {Wada}}]{2014Sci...346..732Z}
{Zemcov}, M., {Smidt}, J., {Arai}, T., {et~al.} 2014, Science, 346, 732

\end{thebibliography}


\appendix
\section{Author Institutions}
\noindent
L.V.E Koopmans,
Kapteyn Astronomical Institute, University of Groningen, The Netherlands\\
J.~Pritchard, Imperial College, U.K.\\
G.~Mellema, Dept. of Astronomy and Oskar Klein Centre, Stockholm Univ., Sweden\\
F.~Abdalla, University College London, U.K.\\
J.~Aguirre, University of Pennsylvania, USA\\
K.~Ahn, Dept. of Earth Science, Chosun University, Korea\\
R. Barkana, Tel Aviv University, Israel \\
I.~van Bemmel, Joint Institute for VLBI in Europe, The Netherlands\\
G.~Bernardi, SKA South Africa \& Rhodes University, South Africa\\
A.~Bonaldi, Jodrell Bank Centre for Astrophysics, University of Manchester, U.K.\\
F.~Briggs, Research School of Astronomy \& Astrophysics, Australian
National University\\
A.G.~de Bruyn, Netherlands Institute for Radio Astronomy (ASTRON), 
The Netherlands\\
T.C.~Chang, Academia Sinica Institute of Astronomy \& Astrophysics, Taiwan\\
E.~Chapman, Department of Physics \& Astronomy, University College London,
U.K.\\
X.~Chen, National Astronomical Observatories,
Chinese Academy of Science, Beijing, China\\
B. Ciardi, Max Planck Institute for Astrophysics, Garching, Germany\\
K.K.~Datta, Dept. of Physics, Presidency University, Kolkata, India\\
P.~Dayal, Institute for Computational Cosmology, Durham University, U.K.\\
A.~Ferrara, Scuola Normale Superiore, Pisa, Italy \\
A.~Fialkov, Ecole Normale Superieure, Paris, France\\
F.~Fiore, INAF Osservatorio Astronomico di Roma, Italy\\
K.~Ichiki, Nagoya University, Japan\\
I.~T.~Iliev, Astronomy Centre, Department of Physics \& Astronomy, 
University of Sussex, U.K.\\
S.~Inoue, Max Planck Institute for Astrophysics, Germany\\
V. Jeli\'{c},
Kapteyn Astronomical Institute, University of Groningen, and ASTRON, The
Netherlands\\
M.~Jones, Oxford University, U.K.\\
J.~Lazio, Jet Propulsion Laboratory, California Institute of Technology,
USA\\
K.~J.~Mack, University of Melbourne, Australia\\
U.~Maio, INAF-Trieste, Italy \& Leibniz Institute for Astrophysics, Germany\\
S.~Majumdar, Stockholm University, Sweden\\
A.~Mesinger, Scuola Normale Superiore, Pisa, Italy\\
M.F.~Morales, Dark Universe Science Center, University of Washington, USA\\
A.~Parsons, University California Berkeley, USA\\
U.-L.~Pen, Canadian Institute for Theoretical Astrophysics, Toronto, Canada\\
M.~Santos, University of Western Cape, South Africa\\
R.~Schneider, INAF/Osservatorio Astronomico di Roma, Italy\\
B.~Semelin, LERMA, Observatoire de Paris, France \\
R.S.~de Souza, MTA Eotvos University, Hungary \\
R.~Subrahmanyan, 
Raman Research Institute, Bangalore 560080, India\\
T.~Takeuchi, Nagoya University, Japan\\
C.~Trott, International Centre for Radio Astronomy Research, Curtin
University, Australia \\
H.~Vedantham, 
Kapteyn Astronomical Institute, University of Groningen, The Netherlands\\
J.~Wagg, Square Kilometre Array Organisation\\
R.~Webster, School of Physics, University of Melbourne, Australia\\
S.~Wyithe, School of Physics, University of Melbourne, Australia

\end{document}